\documentclass[%
reprint,
superscriptaddress,
nofootinbib,
amsmath,amssymb,
aps,
amsthm,
pra,
]{revtex4-1}

\usepackage{graphicx}% Include figure files
\usepackage{dcolumn}% Align table columns on decimal point
\usepackage{bm}% bold math
\usepackage{braket}%
\usepackage{theorem}
\usepackage{multirow}
\usepackage{enumerate}
\usepackage{txfonts}
\usepackage{framed}
\usepackage{bbm}
\usepackage[dvipdfmx,colorlinks,citecolor=blue,linkcolor=blue,urlcolor=blue,bookmarksopenlevel=4]{hyperref}

\usepackage{color}

\definecolor{memo}{RGB}{128,0,255}
\definecolor{gray}{RGB}{128,128,128}

\theorembodyfont{\upshape}

\newtheorem{thm}{Theorem}

\newtheorem{remark}[thm]{Remark}

\newtheorem{cor}[thm]{Corollary}

\newtheorem{proof}{Proof}

\newcommand{\hA}{\hat{A}}
\newcommand{\hB}{\hat{B}}

\newcommand{\hE}{\hat{E}}

\newcommand{\hP}{\hat{P}}

\newcommand{\hS}{\hat{S}}
\newcommand{\hT}{\hat{T}}

\newcommand{\hV}{\hat{V}}

\newcommand{\hX}{\hat{X}}

\newcommand{\hZ}{\hat{Z}}
\newcommand{\ha}{\hat{a}}
\newcommand{\hb}{\hat{b}}
\newcommand{\hc}{\hat{c}}

\newcommand{\hPi}{\hat{\Pi}}

\newcommand{\hPsi}{\hat{\Psi}}

\newcommand{\hGamma}{\hat{\Gamma}}

\newcommand{\hrho}{\hat{\rho}}

\newcommand{\hnu}{\hat{\nu}}

\newcommand{\mC}{\mathcal{C}}
\newcommand{\mD}{\mathcal{D}}

\newcommand{\mH}{\mathcal{H}}
\newcommand{\mI}{\mathcal{I}}

\newcommand{\mK}{\mathcal{K}}
\newcommand{\mL}{\mathcal{L}}
\newcommand{\mM}{\mathcal{M}}

\newcommand{\mT}{\mathcal{T}}

\newcommand{\trho}{\tilde{\rho}}

\newcommand{\PS}{P_{\rm S}}

\newcommand{\ident}{\hat{1}}

\newcommand{\POVM}{\mM}

\newcommand{\QED}{\hspace*{0pt}\hfill $\blacksquare$}

\newcommand{\Tr}{{\rm Tr}}
\newcommand{\rank}{{\rm rank}}
\newcommand{\supp}{{\rm supp}}
\newcommand{\Ker}{{\rm Ker}}

\def\gauss_sym#1{{\lfloor #1 \rfloor}}

\newcommand{\opt}{\star}

\newcommand{\A}{{\rm A}}
\newcommand{\B}{{\rm B}}
\newcommand{\C}{{\rm C}}

\newcommand{\w}{{(\omega)}}
\newcommand{\hPiA}{\hPi^{(\hA)}}
\newcommand{\hBw}{\hB^\w}
\newcommand{\hGopt}{\hGamma^\opt}
\newcommand{\hGone}{\hGamma^{(1)}}
\newcommand{\hGtwo}{\hGamma^{(2)}}

\newcommand{\sumr}{\sum_{r=0}^{R-1}}
\newcommand{\POVMA}{\POVM_\A}
\newcommand{\dw}{d\omega}
\newcommand{\hAdw}{\hA(\dw)}
\newcommand{\TrB}{\Tr_\B}
\newcommand{\mHA}{\mH_\A}
\newcommand{\mHB}{\mH_\B}
\newcommand{\mHC}{\mH_\C}
\newcommand{\mHBC}{\mH_{\rm BC}}
\newcommand{\identA}{\ident_\A}
\newcommand{\identB}{\ident_\B}
\newcommand{\PG}{{\rm P_G}}
\newcommand{\DPG}{{\rm DP_G}}
\newcommand{\G}{{\rm G}}
\newcommand{\pw}{p^\w}
\newcommand{\ew}{e^\w}
\newcommand{\muw}{\mu^\w}
\newcommand{\Tw}{{T^\w}}

\newcommand{\KA}{K_\A}
\newcommand{\KB}{K_\B}
\newcommand{\hZG}{\hZ_\G}
\newcommand{\hXGopt}{\hX_\G^\opt}
\renewcommand{\PS}{P}
\newcommand{\hVA}{\hV_\A}
\newcommand{\hVB}{\hV_\B}
\newcommand{\hVC}{\hV_\C}
\newcommand{\phiA}{\phi}
\newcommand{\phiB}{\phi'}
\newcommand{\tc}{\tilde{c}}
\newcommand{\tu}{\tilde{u}}
\newcommand{\tx}{\tilde{x}}
\newcommand{\tphi}{\tilde{\phi}}
\newcommand{\lift}{{\rm L}}
\newcommand{\SA}{S_\A}
\newcommand{\SB}{S_\B}

\begin{document}

\preprint{APS/123-QED}

\title{Local unambiguous discrimination of symmetric ternary states}

\affiliation{%
 Quantum Information Science Research Center, Quantum ICT Research Institute, Tamagawa University,
 Machida, Tokyo 194-8610, Japan
}%
\affiliation{%
 School of Information Science and Technology,
 Aichi Prefectural University,
 Nagakute, Aichi 480-1198, Japan
}%

\author{Kenji Nakahira}
\affiliation{%
 Quantum Information Science Research Center, Quantum ICT Research Institute, Tamagawa University,
 Machida, Tokyo 194-8610, Japan
}%

\author{Kentaro Kato}
\affiliation{%
 Quantum Information Science Research Center, Quantum ICT Research Institute, Tamagawa University,
 Machida, Tokyo 194-8610, Japan
}%

\author{Tsuyoshi \surname{Sasaki Usuda}}
\affiliation{%
 School of Information Science and Technology,
 Aichi Prefectural University,
 Nagakute, Aichi 480-1198, Japan
}%
\affiliation{%
 Quantum Information Science Research Center, Quantum ICT Research Institute, Tamagawa University,
 Machida, Tokyo 194-8610, Japan
}%

\date{\today}% It is always \today, today,

\begin{abstract}
 We investigate unambiguous discrimination between given quantum states with a sequential
 measurement, which is restricted to local measurements and one-way classical communication.
 %We investigate a sequential (i.e., local operations and one-way classical communication)
 %measurement that unambiguously discriminates between given quantum states.
 If the given states are binary or those each of whose individual systems is two-dimensional,
 then it is in some cases known whether a sequential measurement achieves
 a globally optimal unambiguous measurement.
 In contrast, for more than two states each of whose individual systems is more than two-dimensional,
 the problem becomes extremely complicated.
 This paper focuses on symmetric ternary pure states each of whose individual systems is
 three-dimensional, which include phase shift keyed (PSK) optical coherent states
 and a lifted version of ``double trine'' states.
 We provide a necessary and sufficient condition for
 an optimal sequential measurement to be globally optimal for the bipartite case.
 A sufficient condition of global optimality for multipartite states is also presented.
 One can easily judge whether these conditions hold for given states.
 Some examples are given, which demonstrate that, despite the restriction to
 local measurements and one-way classical communication, a sequential measurement can be
 globally optimal in quite a few cases.
\end{abstract}

% PACS 03.67.Hk: Quantum communication
\maketitle

\section{Introduction}

Discrimination between quantum states as accurately as possible
is a fundamental issue in quantum information theory.
It is a well-known property of quantum theory that
perfect discrimination among nonorthogonal quantum states is impossible.
Then, given a finite set of nonorthogonal quantum states, we need to find an optimal
measurement with respect to a reasonable criterion.
Unambiguous discrimination is one of the most common strategies to
distinguish between quantum states \cite{Iva-1987,Die-1988,Per-1988}.
An unambiguous measurement achieves error-free (i.e., unambiguous) discrimination
at the expense of allowing for a certain rate of inconclusive results.
Finding an unambiguous measurement that maximizes the average success probability
for various quantum states has been widely investigated
(e.g., \cite{Jae-Shi-1995,Ray-Lut-Enk-2003,Eld-Sto-Has-2004,Fen-Dua-Yin-2004,Jaf-Rez-Kar-Ami-2008,Pan-Wu-2009,Kle-Kam-Bru-2010,Sug-Has-Hor-Hay-2010,Ber-Fut-Fel-2012}).

When given quantum states are shared between two or more systems,
measurement strategies can be classified into two types: global and local.
A local measurement is performed by a series of individual measurements on the subsystems
combined with classical communication.
In particular, sequential measurements, in which the classical communication is one-way only,
have been widely investigated under several optimality criteria
(e.g., \cite{Bro-Mei-1996,Ban-Yam-Hir-1997,Vir-Sac-Ple-Mar-2001,Aci-Bag-Bai-Mas-2005,Owa-Hay-2008,Ass-Poz-Pie-2011,Nak-Usu-2012-receiver,Nak-Usu-2016-LOCC,Ros-Mar-Gio-2017-capacity,Cro-Bar-Wei-2017}).
Although the performance of an optimal sequential measurement is often strictly less than
that of an optimal global measurement even if given states are not entangled,
a sequential measurement has the advantage of being relatively easy to implement with current technology.
As an example of a realizable sequential measurement for optical coherent states,
a receiver based on a combination of a photon detector and a feedback circuit,
which we call a Dolinar-like receiver, has been proposed \cite{Dol-1973}
and experimentally demonstrated \cite{Coo-Mar-Ger-2007}.
Also, unambiguous discrimination using Dolinar-like receivers
has been studied \cite{Ban-1999,Enk-2002,Bec-Fan-Mig-2013}.

Several studies on optimal unambiguous sequential measurements have also been carried out
\cite{Che-Yan-2002,Ji-Cao-Yin-2005,Chi-Dua-Hsi-2014,Sen-Mar-Mun-2018}.
For binary pure states with any prior probabilities,
it has been shown that an optimal unambiguous sequential measurement can achieve
the performance of an optimal global measurement \cite{Che-Yan-2002,Ji-Cao-Yin-2005}.
For short, we say that a sequential measurement can be globally optimal.
As for more than two states, in the case in which each of the individual systems is two-dimensional,
whether a sequential measurement can be globally optimal has been clarified for
several cases \cite{Chi-Dua-Hsi-2014,Sen-Mar-Mun-2018}.
However, in the case in which individual systems are more than two-dimensional,
the problem becomes extremely complicated.
Due to the restriction of local measurements and one-way classical communication,
it would not be surprising if a sequential measurement cannot be globally optimal
except for some special cases.
It is worth mentioning that, according to Ref.~\cite{Nak-Kat-Usu-2018-Dolinar},
in the case of a minimum-error measurement, which maximizes the average success probability
but sometimes returns an incorrect answer,
an optimal sequential measurement does not seem to be globally optimal
for any ternary phase shift keyed (PSK) optical coherent states.

In this paper, we focus on symmetric ternary pure states
each of whose individual systems is three-dimensional.
These states include PSK optical coherent states and a lifted version of ``double trine'' states \cite{Sho-2002}.
We provide a necessary and sufficient condition that
a sequential measurement can be globally optimal for the bipartite case,
using which one can easily judge whether global optimality is achieved by a sequential measurement
for given states.
We use the convex optimization approach reported in Ref.~\cite{Nak-Kat-Usu-2018-seq_gen}
to derive the condition.
We also give a sufficient condition of global optimality for the multipartite case.
Some examples of symmetric ternary pure states are presented,
which show that a sequential measurement can be globally optimal in quite a few cases.
One of the examples shows that the problem of whether a sequential measurement
for bipartite ternary PSK optical coherent states can be globally optimal is
completely solved analytically,
while its minimum-error measurement version has been solved only numerically
\cite{Nak-Kat-Usu-2018-Dolinar}.
Moreover, we show that a Dolinar-like receiver for any ternary PSK optical coherent states
cannot be globally optimal unambiguous measurement.

The paper is organized as follows.
In Sec.~\ref{sec:optimization}, we formulate the problem of finding
an optimal unambiguous measurement and its sequential-measurement version
as convex programming problems.
In Sec.~\ref{sec:sym3}, we present our main theorem.
Using this theorem, we derive a necessary and sufficient condition
for an optimal sequential measurement for bipartite symmetric ternary pure states
to be globally optimal.
A sufficient condition of global optimality for multipartite
symmetric ternary pure states is also derived.
In Sec.~\ref{sec:sym3_optLambda}, we prove the main theorem.
Finally, we provide some examples to demonstrate the usefulness of our results
in Sec.~\ref{sec:example}.

\section{Optimal unambiguous sequential measurements} \label{sec:optimization}

In this section, we first provide an optimization problem of finding optimal
unambiguous measurements.
Then, we discuss a sequential-measurement version of the optimization problem.
We also provide a necessary and sufficient condition for an optimal sequential measurement
to be globally optimal.
Note that this condition is quite general but requires extra effort to decide
whether global optimality is achieved by a sequential measurement for given quantum states.
In Sec.~\ref{sec:sym3}, we will use this condition to derive a formula that is directly applicable to
symmetric ternary pure states.

\subsection{Problem of finding optimal unambiguous measurements}

We here consider unambiguous measurements without restriction to sequential measurements.
%# if we have the restriction of xx
%# under the restriction to xx
Consider a quantum system prepared in one of $R$ quantum states represented by density operators
$\{ \trho_r \}_{r \in \mI_R}$ on a complex Hilbert space $\mH$, where $\mI_R \coloneqq \{ 0, 1, \cdots, R-1 \}$.
The density operator $\trho_r$ satisfies $\trho_r \ge 0$ and $\Tr ~\trho_r = 1$,
where $\hA \ge 0$ denotes that $\hA$ is positive semidefinite
(similarly, $\hA \ge \hB$ denotes $\hA - \hB \ge 0$).
To unambiguously discriminate the $R$ states, we can
consider a measurement represented by
a positive-operator-valued measure (POVM),
$\hPi \coloneqq \{ \hPi_r \}_{r=0}^R$,
consisting of $R + 1$ detection operators, on $\mH$,
where $\hPi_r$ satisfies $\hPi_r \ge 0$ and $\sum_{r=0}^R \hPi_r = \ident$
($\ident$ is the identity operator on $\mH$).
The detection operator $\hPi_r$ with $r < R$ corresponds to the identification of the
state $\trho_r$, while $\hPi_R$ corresponds to the inconclusive answer.
%(``I don't know'').
Any unambiguous measurement $\hPi$ satisfies $\Tr(\trho_r \hPi_k) = 0$ for any $k \in \mI_R \backslash \{r\}$,
where $\backslash$ denotes set difference.
Given possible states $\{ \trho_r \}$ and their prior probabilities $\{ \xi_r \}$,
we want to find an unambiguous measurement maximizing the average success probability,
which we call an optimal unambiguous measurement or just an optimal measurement for short.
Reference~\cite{Eld-Sto-Has-2004} shows that
the problem of finding an optimal measurement can be formulated as a semidefinite programming problem,
which is a special case of a convex programming problem.
For analytical convenience, instead of the formulation of Ref.~\cite{Eld-Sto-Has-2004},
we consider the following semidefinite programming problem:
\begin{eqnarray}
 \begin{array}{lll}
  {\rm \PG:} & {\rm maximize} & \displaystyle \PS(\hPi) \coloneqq
   \lim_{\lambda \to \infty} \sumr \Tr [ (\hrho_r - \lambda \hnu_r) \hPi_r ] \\
  & {\rm subject~to} & \hPi : {\rm POVM},
 \end{array} \label{eq:PG}
\end{eqnarray}
where $\hrho_r \coloneqq \xi_r \trho_r$ and $\hnu_r \coloneqq \sum_{k \in \mI_R \backslash \{ r \}} \hrho_k$.
Since $\PS(\hPi) = -\infty$ holds if there exists $r \in \mI_R$ such that
$\Tr(\hnu_r \hPi_r) \neq 0$ (i.e., $\hPi$ is not an unambiguous measurement),
any optimal solution to Problem~$\PG$ is guaranteed to be an unambiguous measurement.
The optimal value, which is the average success probability of an optimal measurement,
is larger than zero if and only if at least one of the operators $\hrho_r$ has a nonzero overlap
with the kernels of $\hnu_r$ \cite{Rud-Spe-Tur-2003}.

The dual problem to Problem~$\PG$ can be written as
\footnote{One can obtain this problem from Eq.~(12) in Ref.~\cite{Nak-Kat-Usu-2015-general}
with $M = R + 1$, $J = 0$, $\hc_m = \hrho_m - \lambda \hnu_m$ $~(m < R)$, $\hc_R = 0$,
and $\lambda \to \infty$.}
\begin{eqnarray}
 \begin{array}{lll}
  {\rm \DPG:} & {\rm minimize} & \Tr~\hZG \\
  & {\rm subject~to} & \displaystyle \hZG \ge
   \lim_{\lambda \to \infty} \hrho_r - \lambda \hnu_r ~ (\forall r \in \mI_R),
 \end{array} \label{eq:DPG}
\end{eqnarray}
where $\hZG$ is a positive semidefinite operator on $\mH$.
The optimal values of Problems~$\PG$ and $\DPG$ are the same.

\subsection{Problem of finding optimal unambiguous sequential measurements}

Now, let us assume that $\mH$ is a bipartite Hilbert space, $\mH = \mHA \otimes \mHB$,
and let us restrict our attention to a sequential measurement from Alice to Bob.
In a sequential measurement, Alice performs a measurement on $\mHA$ and
communicates her result to Bob.
Then, he performs a measurement on $\mHB$, which can depend on Alice's outcomes,
and obtains the final measurement result.
This sequential measurement can be considered from a different point of view
\cite{Nak-Kat-Usu-2018-Dolinar}.
Let $\omega$ be an index associated with Bob's measurement $\hBw \coloneqq \{ \hBw_r \}_{r=0}^R$,
and $\Omega$ be the entire set of indices $\omega$.
Alice performs a measurement, $\hA \coloneqq \{ \hA(\omega) \}_{\omega \in \Omega}$,
with continuous outcomes, and sends the result $\omega \in \Omega$ to Bob.
Then, he performs the corresponding measurement $\hBw$,
which is uniquely determined by the result $\omega$.
This sequential measurement is denoted as $\hPiA \coloneqq \{ \hPiA_r \}_{r=0}^R$ with
\begin{eqnarray}
 \hPiA_r &\coloneqq& \int_{\Omega} \hAdw \otimes \hBw_r, \label{eq:PiA}
\end{eqnarray}
which is uniquely determined by Alice's POVM $\hA$.

The problem of finding an unambiguous sequential measurement maximizing the average success probability,
which we call an optimal unambiguous sequential measurement or just an optimal sequential measurement,
can be formulated as the following optimization problem:
\begin{eqnarray}
 \begin{array}{lll}
  {\rm P:} & {\rm maximize} & \displaystyle \PS[\hPiA] \\
  & {\rm subject~to} & \hA \in \POVMA
 \end{array} \label{eq:P}
\end{eqnarray}
with Alice's POVM $\hA$,
where $\POVMA$ is the entire set of Alice's continuous measurements $\{ \hA(\omega) \}_{\omega \in \Omega}$.
Compared to Problem~$\PG$, this problem restricts $\hPi$ to the form $\hPi = \hPiA$.
We can easily see that this problem is a convex programming problem
and obtain the following dual problem \cite{Nak-Kat-Usu-2018-seq_gen}:
\begin{eqnarray}
 \begin{array}{lll}
  {\rm DP:} & {\rm minimize} & \Tr~\hX \\
  & {\rm subject~to} & \displaystyle \hGamma(\omega; \hX) \ge 0 ~ (\forall \omega \in \Omega)
 \end{array} \label{eq:DP}
\end{eqnarray}
with a Hermitian operator $\hX$, where
\begin{eqnarray}
 \hGamma(\omega; \hX) &\coloneqq&
  \hX - \lim_{\lambda \to \infty} \sumr \TrB \left[ (\hrho_r - \lambda \hnu_r) \hBw_r \right].
  \label{eq:hGw}
\end{eqnarray}
$\TrB$ is the partial trace over $\mHB$.
The optimal values of Problems~P and DP are also the same.

\subsection{Condition for sequential measurement to be globally optimal}

Let $\hZG^\opt$ be an optimal solution to Problem~$\PG$ and $\hXGopt \coloneqq \TrB~\hZG^\opt$.
Also, let $\hGopt(\omega) \coloneqq \hGamma(\omega; \hXGopt)$.
We now want to know whether a sequential measurement can be globally optimal,
i.e., whether an optimal solution to Problem~P is also optimal to Problem~$\PG$.
To this end, we utilize the following remark:
\begin{remark} \label{remark:opt}
 A sequential measurement $\hPiA$ $~(\hA \in \POVMA)$ is an optimal unambiguous measurement
 if and only if it satisfies
 \begin{eqnarray}
  \hGopt(\omega) \hA(\omega) &=& 0, ~~ \forall \omega \in \Omega. \label{eq:cond}
 \end{eqnarray}
\end{remark}

\begin{proof}
 Assume that $\hXGopt$ is a feasible solution to Problem~DP,
 i.e., $\hGopt(\omega) \ge 0$ holds for any $\omega \in \Omega$.
 It is known that $\hPiA$ and $\hX$ are respectively optimal solutions to Problems~P and DP
 if and only if $\hGamma(\omega; \hX) \ge 0$ and $\hGamma(\omega; \hX) \hA(\omega) = 0$ hold
 for any $\omega \in \Omega$
 (see Theorem~2 of Ref.~\cite{Nak-Kat-Usu-2018-seq_gen}
 \footnote{We here consider the case of $M = R + 1$, $J = 0$, $\hc_m = \hrho_m - \lambda \hnu_m$ $~(m < R)$,
 $\hc_R = 0$, and $\lambda \to \infty$.}).
 Thus, $\hPiA$ and $\hXGopt$ are respectively optimal solutions to Problems~P and DP
 if and only if Eq.~\eqref{eq:cond} holds.
 If Eq.~\eqref{eq:cond} holds, then, since $P[\hPiA] = \Tr~\hXGopt = \Tr~\hZG^\opt$ is equal to
 the optimal value of Problem~$\PG$, $\hPiA$ is globally optimal.
 Therefore, to prove this remark, it suffices to show that $\hXGopt$ is a feasible solution to Problem~DP.

 Multiplying $[ \hBw_r ]^{1/2}$ on both sides of the constraint of Problem~$\DPG$
 and taking the partial trace over $\mHB$ gives
 \begin{eqnarray}
  \TrB \left[ \hZG^\opt \hBw_r \right] &\ge& \lim_{\lambda \to \infty}
   \TrB \left[ (\hrho_r - \lambda \hnu_r) \hBw_r \right].
 \end{eqnarray}
 Therefore, we have
 \begin{eqnarray}
  \sumr \TrB \left[ \hZG^\opt \hBw_r \right] &\ge& \lim_{\lambda \to \infty}
   \sumr \TrB \left[ (\hrho_r - \lambda \hnu_r) \hBw_r \right].
 \end{eqnarray}
 Also, from $\hXGopt = \TrB~\hZG^\opt$, we have
 \begin{eqnarray}
  \hXGopt &=& \sum_{r=0}^R \TrB \left[ \hZG^\opt \hBw_r \right]
   \ge \sumr \TrB \left[ \hZG^\opt \hBw_r \right].
 \end{eqnarray}
 From these equations and Eq.~\eqref{eq:hGw}, $\hGopt(\omega) \ge 0$ holds for any $\omega \in \Omega$,
 and thus $\hXGopt$ is a feasible solution to Problem~DP.
 \QED
\end{proof}

We will further investigate Alice's POVM $\hA$ satisfying Eq.~\eqref{eq:cond}.
Let
\begin{eqnarray}
 \mK_\omega &\coloneqq& \Ker~\sumr \TrB \left[ \hnu_r \hBw_r \right]. \label{eq:mK_omega}
\end{eqnarray}
Let us consider $\ket{\gamma} \in \supp~\hA(\omega)$.
Suppose that Eq.~\eqref{eq:cond} holds;
then, from Eqs.~\eqref{eq:hGw} and \eqref{eq:mK_omega},
$\ket{\gamma} \in \mK_\omega$ and
$\hP_\omega \left[ \hXGopt - \sumr \TrB [\hrho_r \hBw_r] \right] \ket{\gamma} = 0$ hold,
where $\hP_\omega$ is the projection operator onto $\mK_\omega$.
Conversely, if these two equations hold for any $\ket{\gamma} \in \supp~\hA(\omega)$,
then Eq.~\eqref{eq:cond} holds.
Therefore, Eq.~\eqref{eq:cond} is equivalent to the following equations:
\begin{eqnarray}
 \supp~\hA(\omega) &\subseteq& \mK_\omega, \nonumber \\
 \hP_\omega \left[ \hXGopt - \sumr \TrB \left[ \hrho_r \hBw_r \right] \right] \hA(\omega) &=& 0.
  \label{eq:cond2}
\end{eqnarray}

Let us consider the case in which each state $\hrho_r$ is separable,
i.e., it is in the form of
\begin{eqnarray}
 \hrho_r &=& \xi_r \ha_r \otimes \hb_r, \label{eq:separable}
\end{eqnarray}
where $\ha_r$ and $\hb_r$ are respectively density operators on $\mHA$ and $\mHB$.
Then, Eq.~\eqref{eq:hGw} reduces to
\begin{eqnarray}
 \hGamma(\omega; \hX) &=& \hX - \sumr \pw_r \xi_r \ha_r
  + \lim_{\lambda \to \infty} \lambda \sum_{r=0}^{R-1} \ew_r \xi_r \ha_r, \label{eq:hGw2}
\end{eqnarray}
where $\pw_r \coloneqq \Tr [ \hb_r \hBw_r ]$ is
the probability of Bob correctly identifying the state $\hb_r$
and $\ew_r \coloneqq \sum_{k \in \mI_R \backslash \{ r \}} \Tr [ \hb_r \hBw_k ]$
is the probability of Bob misidentifying the state $\hb_r$.
Also, it follows from $\mK_\omega = \Ker~\sum_{r=0}^{R-1} \ew_r \xi_r \ha_r$ that
the first line of Eq.~\eqref{eq:cond2} can be expressed as
\begin{eqnarray}
 \Tr[\ha_r \hA(\omega)] &=& 0, ~~ \forall r \not\in \Tw, \label{eq:suppA}
\end{eqnarray}
where $\Tw$ is the entire set of indices $r \in \mI_R$ such that
Bob's measurement never gives incorrect results, i.e.,
\begin{eqnarray} 
 \Tw &\coloneqq& \left\{ r \in \mI_R : \ew_r = 0 \right\}. \label{eq:Tw}
\end{eqnarray}
%
% which we call the indices unambiguously identified by $\hBw$.
Equation~\eqref{eq:suppA} implies that, for any $r \in \mI_R$ and $\omega \in \Omega$ such that
the state $\hb_r$ will be incorrectly identified by Bob's measurement $\hBw$
(i.e., $\ew_r \neq 0$),
Alice's outcome must not be $\omega$ for the state $\ha_r$ (i.e., $\Tr[\ha_r \hA(\omega)] = 0$).
Thus, Eq.~\eqref{eq:suppA} ensures that the measurement $\hPiA$ never gives erroneous results.

\section{Sequential measurements for symmetric ternary pure states} \label{sec:sym3}

Remark~\ref{remark:opt} is useful in determining whether
a sequential measurement can be globally optimal.
Concretely, it is possible to decide whether a sequential measurement can be globally optimal
by examining whether there exists $\hA \in \POVMA$ satisfying Eq.~\eqref{eq:cond}.
However, in general, it is quite difficult to examine this for all continuous values $\omega \in \Omega$.
In this section, we consider sequential measurements for bipartite symmetric ternary pure states
and derive a formula that can directly determine whether a sequential measurement can be globally optimal.
Extending our results to the multipartite case enables us to obtain
a sufficient condition that a sequential measurement can be globally optimal.

\subsection{Main results}

Let us consider bipartite ternary separable pure states,
$\{ \ket{\Psi_r} \coloneqq \ket{a_r} \otimes \ket{b_r} \}_{r=0}^2$,
which are the special case of Eq.~\eqref{eq:separable} with
$\ha_r = \ket{a_r}\bra{a_r}$ and $\hb_r = \ket{b_r}\bra{b_r}$.
Assume that $\{ \ket{a_r} \}$ and $\{ \ket{b_r} \}$ respectively span
three-dimensional Hilbert spaces, $\mHA$ and $\mHB$.
Also, assume that $\{ \ket{\Psi_r} \}$ is symmetric
in the following sense:
the prior probabilities are equal (i.e., $\xi_r = 1/3$) and
there exist unitary operators $\hVA$ on $\mHA$ and $\hVB$ on $\mHB$ satisfying
\begin{eqnarray}
 \ket{a_{r \oplus 1}} &=& \hVA \ket{a_r},
  ~ \ket{b_{r \oplus 1}} = \hVB \ket{b_r}, \label{eq:Vab}
\end{eqnarray}
where $\oplus$ denotes addition modulo 3.
These states are characterized by the inner products
$\KA \coloneqq \braket{a_0|a_1}$ and $\KB \coloneqq \braket{b_0|b_1}$,
which are generally complex values.
For any $r \in \mI_3$, we have
\begin{eqnarray}
 \braket{a_r | a_{r \oplus 1}} &=& \KA, ~~~ \braket{b_r | b_{r \oplus 1}} = \KB.
\end{eqnarray}
${ \ket{a_r} }$ and/or ${ \ket{b_r} }$ can be 
PSK optical coherent states,
pulse position modulated (PPM) optical coherent states,
and lifted trine states \cite{Sho-2002}.
If $\{ \ket{a_r} \}$ or $\{ \ket{b_r} \}$ is mutually orthogonal (i.e., $\KA = 0$ or $\KB = 0$),
then an optimal sequential measurement perfectly discriminates $\{ \ket{\Psi_r} \}$,
and thus is globally optimal.
So, assume that $\{ \ket{a_r} \}$ and $\{ \ket{b_r} \}$ are not mutually orthogonal.

We shall present a theorem that can be used to determine whether
a sequential measurement can be globally optimal for given bipartite symmetric ternary pure states.
Let us consider the following set with seven elements
\begin{eqnarray}
 \Omega^\opt &\coloneqq& \{ \omega_{1,j}, \omega_{2,j}, \omega_3 : j \in \mI_3 \},
  \label{eq:Omega_opt}
\end{eqnarray}
where
\begin{enumerate}[(1)]
 \item $\hB^{(\omega_{1,j})}$ is the measurement that always returns $j$,
       i.e., $\hB^{(\omega_{1,j})}_r = \delta_{r,j} \identB$,
       where $\delta_{r,j}$ is the Kronecker delta and $\identB$ is the identity operator on $\mHB$.
 \item $\hB^{(\omega_{2,j})}$ is an optimal unambiguous measurement
       for binary states $\{ \ket{b_{j \oplus 1}}, \ket{b_{j \oplus 2}} \}$
       with equal prior probabilities of $1/2$.
 \item $\hB^{(\omega_3)}$ is an optimal unambiguous measurement
       for ternary states $\{ \ket{b_r} \}_{r=0}^2$ with equal prior probabilities of $1/3$.
\end{enumerate}
For simpler notation, we write $\omega_k$ for $\omega_{k,0}$ for each $k \in \{1,2\}$.

\begin{figure}[tb]
 \centering
 \includegraphics[scale=0.8]{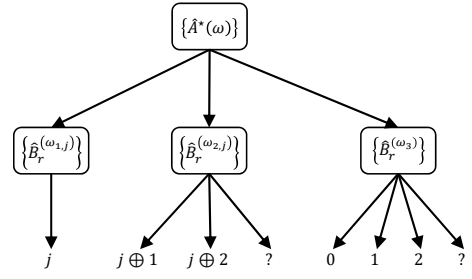}
 \caption{Schematic diagram of an optimal sequential measurement $\hPi^{(\hA^\opt)}$.}
 \label{fig:optPOVM}
\end{figure}

When a sequential measurement can be globally optimal,
there can exist a large (or even infinite) number of optimal sequential measurements.
However, as we shall show in the following theorem,
if a sequential measurement can be globally optimal,
then there always exists an optimal sequential measurement
in which Alice never returns an index $\omega$ with $\omega \not\in \Omega^\opt$
(proof in Sec.~\ref{sec:sym3_optLambda}):
\begin{thm} \label{thm:sym3_optLambda}
 Suppose that, for bipartite symmetric ternary pure states
 $\{ \ket{\Psi_r} \coloneqq \ket{a_r} \otimes \ket{b_r} \}_{r=0}^2$,
 a sequential measurement can be globally optimal.
 Then, there exists an optimal sequential measurement $\hPi^{(\hA^\opt)}$ with $\hA^\opt \in \POVMA$
 such that
 \begin{eqnarray}
  \hA^\opt(\omega) &=& 0, ~~ \forall \omega \not\in \Omega^\opt.
   \label{eq:sym3_optLambda}
 \end{eqnarray}
\end{thm}
The measurement $\hPi^{(\hA^\opt)}$ is schematically illustrated in Fig.~\ref{fig:optPOVM}.
Due to the definition of $\omega_{1,j}, \omega_{2,j}, \omega_3 \in \Omega^\opt$,
$\Tw$, defined by Eq.~\eqref{eq:Tw}, satisfies
$T^{(\omega_{1,j})} = \{ j \}$, $T^{(\omega_{2,j})} = \{ j \oplus 1, j \oplus 2 \}$,
and $T^{(\omega_3)} = \{ 0, 1, 2 \}$.
From Eq.~\eqref{eq:suppA}, $\Tr[\ha_r \hA^\opt(\omega_{1,j})] = 0$ must hold
for any distinct $r,j \in \mI_3$.
Thus, if Alice returns the index $\omega_{1,j}$,
then the given state must be $\ket{\Psi_j}$.
(In this case, the given state is uniquely determined before Bob performs the measurement.)
Also, from Eq.~\eqref{eq:suppA}, $\Tr[\ha_j \hA^\opt(\omega_{2,j})] = 0$ holds for any $j \in \mI_3$,
which indicates that if Alice returns the index $\omega_{2,j}$,
then the state $\ket{\Psi_j}$ is unambiguously filtered out.
In this case, Alice's measurement result does not indicate
which of the two states $\ket{\Psi_{j \oplus 1}}$ and $\ket{\Psi_{j \oplus 2}}$ is given.
If Alice returns the index $\omega_3$, then
Alice's result provides no information about the given state.

Using Theorem~\ref{thm:sym3_optLambda},
we can derive a simple formula for determining whether
a sequential measurement can be globally optimal.
Before we state this formula, we shall give some preliminaries.
Let $\tau \coloneqq \exp(i 2\pi/3)$, where $i \coloneqq \sqrt{-1}$.
Also, let $\ket{\phiA_n}$ and $\ket{\phiB_n}$, respectively, denote
the normalized eigenvectors corresponding to the eigenvalues $\tau^n$ $~(n \in \mI_3)$
of $\hVA$ and $\hVB$.
Moreover, let
\begin{eqnarray}
 x_n &\coloneqq& |\braket{\phiA_n | a_0}|, ~~
  y_n \coloneqq |\braket{\phiB_n | b_0}|.
\end{eqnarray}
Note that $x_n, y_n > 0$ holds for any $n \in \mI_3$.
By selecting appropriate global phases of $\ket{a_r}$ and $\ket{b_r}$ and
permuting $\ket{\Psi_1}$ and $\ket{\Psi_2}$ if necessary,
we may assume
\begin{eqnarray}
 x_0 &>& x_2, ~~ x_1 \ge x_2, ~~~ y_0 \ge y_1 \ge y_2, ~~ y_0 \neq y_2. \label{eq:xy_cond}
\end{eqnarray}
Also, by selecting global phases of $\ket{\phiA_n}$ and $\ket{\phiB_n}$
such that $\braket{\phiA_n | a_0}$ and $\braket{\phiB_n | b_0}$ are positive real numbers,
$\ket{a_r}$ and $\ket{b_r}$ are written as
\begin{eqnarray}
 \ket{a_r} &=& \sum_{n=0}^2 x_n \tau^{rn} \ket{\phiA_n}, ~
 \ket{b_r} = \sum_{n=0}^2 y_n \tau^{rn} \ket{\phiB_n}.
 \label{eq:ab_matrix}
\end{eqnarray}
$x_n$ and $y_n$ are uniquely determined by $\KA$ and $\KB$.
Let $\KA' \coloneqq \braket{a_0|a_1}$ and $\KB' \coloneqq \braket{b_0|b_1}$,
where $\ket{a_r}$ and $\ket{b_r}$ are expressed by Eq.~\eqref{eq:xy_cond} and \eqref{eq:ab_matrix};
then, we have
\begin{eqnarray}
 x_n &=& \sqrt{\frac{1 + \tau^{2n} \KA' + \tau^n (\KA')^*}{3}}, ~~
  y_n = \sqrt{\frac{1 + \tau^{2n} \KB' + \tau^n (\KB')^*}{3}}, \nonumber \\
 \label{eq:xy}
\end{eqnarray}
where $^*$ designates complex conjugate.
Let $\eta \coloneqq (1 - |\KB|) / 3$.
Note that $3\eta = 1 - |\KB|$ equals the average success probability of
the optimal unambiguous measurement $\hB^{(\omega_2)}$
for binary states $\{ \ket{b_1}, \ket{b_2} \}$
with equal prior probabilities of $1/2$.

We get the following corollary (proof in Appendix~\ref{append:cor_sym3_nas}):
\begin{cor} \label{cor:sym3_nas}
 For bipartite symmetric ternary pure states
 $\{ \ket{\Psi_r} \coloneqq \ket{a_r} \otimes \ket{b_r} \}_{r=0}^2$
 expressed by Eqs.~\eqref{eq:xy_cond} and \eqref{eq:ab_matrix},
 the following two statements are equivalent.
 \begin{enumerate}[(1)]
  \item A sequential measurement can be globally optimal.
  \item Either $y_1 = y_2$ or
        \begin{eqnarray}
         x_2 z_0 - x_1 z_1 &\ge& 0, \nonumber \\
         \sum_{k=0}^2 x_k^2 ( z_{1 \ominus k}^{-2} - z_{3 \ominus k}^{-2} ) &\ge& 0
          \label{eq:sym3_nas_cond}
        \end{eqnarray}
        holds, where $z_k \coloneqq y_k^2 - \eta$ and
        $\ominus$ denotes subtraction modulo 3.
 \end{enumerate}
\end{cor}
Using this corollary, we can easily judge whether
a sequential measurement can be globally optimal for 
bipartite symmetric ternary pure states.

\subsection{Extension to multipartite states}

We can extend the above results to multipartite states.
As a simple example, we consider tripartite symmetric ternary pure states
$\{ \ket{\Psi_r} = \ket{a_r} \otimes \ket{b_r} \otimes \ket{c_r} \}_{r=0}^2$,
which have equal prior probabilities.
There exist unitary operators $\hVA$, $\hVB$, and $\hVC$ on $\mHA$, $\mHB$, and $\mHC$
satisfying Eq.~\eqref{eq:Vab} and $\ket{c_{r \oplus 1}} = \hVC \ket{c_r}$.
Here, let us consider the composite system of $\mHB$ and $\mHC$, $\mHBC \coloneqq \mHB \otimes \mHC$,
and interpret these states as bipartite states $\{ \ket{\Psi_r} = \ket{a_r} \otimes \ket{B_r} \}_{r=0}^2$,
where $\ket{B_r} \coloneqq \ket{b_r} \otimes \ket{c_r} \in \mHBC$.
It is obvious that if a sequential measurement can be globally optimal for the tripartite states,
then it is also true for the bipartite states.
Assume that it is true for the bipartite states;
then, from Theorem~\ref{thm:sym3_optLambda}, there exists a sequential measurement $\hPi^{(\hA^\opt)}$
satisfying Eq.~\eqref{eq:sym3_optLambda}, which is globally optimal.
Also, it follows that $\hPi^{(\hA^\opt)}$ can be realized by a sequential measurement on
the tripartite system $\mHA \otimes \mHB \otimes \mHC$
if and only if, for any $\omega \in \Omega^\opt$, the measurement $\hBw$ on $\mHBC$ can be realized
by a sequential measurement on $\mHB \otimes \mHC$.
$\hB^{(\omega_{1,j})} = \{ \hB^{(\omega_{1,j})}_r = \delta_{r,j} \}_r$ can obviously be realized
by a sequential measurement.
Also, it is known that a globally optimal measurement for any bipartite binary pure states
can be realized by a sequential measurement \cite{Che-Yan-2002,Ji-Cao-Yin-2005},
and thus $\hB^{(\omega_{2,j})}$ can also be realized by a sequential one.
Therefore, $\hPi^{(\hA^\opt)}$ can be realized by a sequential measurement on the tripartite system
if and only if $\hB^{(\omega_3)}$ can be realized by a sequential measurement.
Since $\hB^{(\omega_3)}$ is globally optimal for
the bipartite symmetric ternary pure states $\{ \ket{b_r} \otimes \ket{c_r} \}_{r=0}^2$,
$\hB^{(\omega_3)}$ can be realized by a sequential measurement
if and only if a sequential measurement for $\{ \ket{b_r} \otimes \ket{c_r} \}_r$
can be globally optimal.
We can summarize the above discussion as follows:
if a sequential measurement can be globally optimal
for each of the two sets of states $\{ \ket{a_r} \otimes \ket{B_r} \}_r$ and $\{ \ket{b_r} \otimes \ket{c_r} \}_r$,
then the same is true for the tripartite states $\{ \ket{\Psi_r} \}$.

Repeating the above arguments, we can extend it to more than three-partite system,
as stated in the following corollary:
\begin{cor} \label{cor:multipartite}
 Let us consider $N$-partite ternary pure states
 $\{ \ket{\Psi_r} \coloneqq \ket{\psi^{(0)}_r} \otimes \ket{\psi^{(1)}_r} \otimes \cdots
 \otimes \ket{\psi^{(N-1)}_r} \}_{r=0}^2$ with equal prior probabilities,
 where $N \ge 3$.
 Suppose that $\{ \ket{\Psi_r} \}$ are symmetric, i.e.,
 for any $n \in \mI_N$, there exists a unitary operator $\hV^{(n)}$
 satisfying $\ket{\psi^{(n)}_{r \oplus 1}} = \hV^{(n)} \ket{\psi^{(n)}_r}$.
 Let $\ket{b^{(n)}_r} \coloneqq \ket{\psi^{(n+1)}_r} \otimes \cdots \otimes \ket{\psi^{(N-1)}_r}$
 $~(n \in \mI_{N-1})$.
 If for any $n \in \mI_{N-1}$,
 a sequential measurement can be globally optimal for bipartite states
 $\{ \ket{\psi^{(n)}_r} \otimes \ket{b^{(n)}_r} \}_{r=0}^2$
 with equal prior probabilities,
 then the same is true for $\{ \ket{\Psi_r} \}$.
\end{cor}
By using Corollary~\ref{cor:sym3_nas}, one can easily judge whether a sequential measurement
for the bipartite states $\{ \ket{\psi^{(n)}_r} \otimes \ket{b^{(n)}_r} \}_{r=0}^2$ can be globally optimal.
Note that the above sufficient condition may not be necessary.
For example, let us again consider the tripartite states
$\{ \ket{a_r} \otimes \ket{b_r} \otimes \ket{c_r} \}_r$.
For an optimal sequential measurement for these states to be globally optimal,
it is sufficient that there exists a globally optimal sequential measurement
$\hPiA$ for the bipartite states $\{ \ket{a_r} \otimes \ket{B_r} \}_r$
such that the measurement $\hBw$ can be realized by
a sequential measurement on the bipartite system $\mHB \otimes \mHC$
for any $\omega$ with $\hA(\omega) \neq 0$,
where $A$ can be different from $\hA^\opt$.

\section{Proof of Theorem~\ref{thm:sym3_optLambda}} \label{sec:sym3_optLambda}

We now prove Theorem~\ref{thm:sym3_optLambda} using Remark~\ref{remark:opt}.
We provide an overview of our proof in this section, leaving some technical details to the appendix.
As a starting point, we first obtain $\hXGopt$ in Sec.~\ref{subsec:sym3_optLambda_XG}.
Next, in Sec.~\ref{subsec:sym3_optLambda_y1y2}, we consider the case $y_1 = y_2$.
After that, we consider the case $y_1 \neq y_2$.
A sufficient condition for this theorem to hold and its reformulation
are given in Secs.~\ref{subsec:sym3_optLambda_sufficient1} and
\ref{subsec:sym3_optLambda_sufficient2}, respectively.
In Sec.~\ref{subsec:sym3_optLambda_proof}, we prove that this sufficient condition holds.

\subsection{Derivation of $\hXGopt$} \label{subsec:sym3_optLambda_XG}

Let us consider $\ket{a_r}$ and $\ket{b_r}$ in the form of
Eqs.~\eqref{eq:xy_cond} and \eqref{eq:ab_matrix}.
A simple calculation gives
\begin{eqnarray}
 \ket{\Psi_r} &\coloneqq& \ket{a_r} \otimes \ket{b_r} = \sum_{n=0}^2 \tx_n \tau^{rn} \ket{\tphi_n}, \label{eq:Psi}
\end{eqnarray}
where
\begin{eqnarray}
 \ket{\tphi_n} &\coloneqq& \frac{1}{\tx_n} \sum_{k=0}^2 x_k y_{n \ominus k}
  \ket{\phiA_k} \otimes \ket{\phiB_{n \ominus k}}, \nonumber \\
 \tx_n &\coloneqq& \sqrt{\sum_{k=0}^2 x_k^2 y_{n \ominus k}^2}.
\end{eqnarray}
Obviously, $\{ \ket{\tphi_n} \}_{n=0}^2$ is an orthonormal basis
and $\tx_n$ is positive real.
We have
\begin{eqnarray}
 \tx_1^2 - \tx_2^2 &=& x_0y_1 + x_1y_0 + x_2y_2 - x_0y_2 - x_1y_1 - x_2y_0 \nonumber \\
 &=& (x_0 - x_2)(y_1 - y_2) + (x_1 - x_2)(y_0 - y_1) \nonumber \\
 &\ge& 0, \label{eq:tx12}
\end{eqnarray}
where the inequality follows from Eq.~\eqref{eq:xy_cond}.
Thus, $\tx_1 \ge \tx_2$ holds.
Note that whether $\tx_0 \ge \tx_2$ or not depends on given states.
Let
\begin{eqnarray}
 [\upsilon_0, \upsilon_1, \upsilon_2] &\coloneqq&
  \left\{
   \begin{array}{cc}
    \left[ 2, 1, 0 \right], & \tx_0 \ge \tx_2, \\
    \left[ 0, 2, 1 \right], & {\rm otherwise}; \\
   \end{array}
  \right. \label{eq:upsilon}
\end{eqnarray}
then, $\tx_{\upsilon_1} \ge \tx_{\upsilon_0}$ and $\tx_{\upsilon_2} \ge \tx_{\upsilon_0}$ hold.
We can easily see that an optimal solution to Problem~$\DPG$ is given by
(see Appendix~\ref{append:sym3_opt})
\begin{eqnarray}
 \hZG^\opt &=& 3\tx_{\upsilon_0}^2 \ket{\tphi_{\upsilon_0}} \bra{\tphi_{\upsilon_0}}. \label{eq:ZG}
\end{eqnarray}
From Eq.~\eqref{eq:ZG}, we have
\begin{eqnarray}
 \hXGopt &=& \TrB~\hZG^\opt = 3 \sum_{n=0}^2 x_n^2 y_{\upsilon_n}^2 \ket{\phiA_n} \bra{\phiA_n}. \label{eq:XG}
\end{eqnarray}

\subsection{Case of $y_1 = y_2$} \label{subsec:sym3_optLambda_y1y2}

We here show that, in the case of $y_1 = y_2$ (i.e., $\KB$ is positive real),
there exists a globally optimal sequential measurement $\hPi^{(\hA^\opt)}$
satisfying Eq.~\eqref{eq:sym3_optLambda}.
Let
\begin{eqnarray}
 \hA^\opt(\omega) &\coloneqq&
  \left\{
   \begin{array}{cc}
    \hA_r, & \omega = \omega_{1,r} ~ (r \in \mI_3), \\
    \hA_3, & \omega = \omega_3, \\
    0, & {\rm otherwise}, \\
   \end{array}
  \right.
\end{eqnarray}
where $\{ \hA_r \}_{r=0}^3$ is an optimal unambiguous measurement for $\{ \ket{a_r} \}$
with equal prior probabilities.
Obviously, $\hA^\opt$ is in $\POVMA$ and satisfies Eq.~\eqref{eq:sym3_optLambda}.
It follows that the sequential measurement $\hPi^{(\hA^\opt)}$ can be interpreted as follows:
Alice and Bob respectively perform optimal measurements for $\{ \ket{a_r} \}$ and $\{ \ket{b_r} \}$
with equal prior probabilities and get the results, $r_\A$ and $r_\B$.
$\hPi^{(\hA^\opt)}$ returns $r_\A$ if $r_\A \in \mI_3$, $r_\B$ if $r_\B \in \mI_3$, and
$r = 3$ otherwise.
Note that $r_\A = r_\B$ holds whenever $r_\A$ and $r_\B$ are in $\mI_3$.

The average success probabilities of optimal measurements for $\{ \ket{a_r} \}$ and $\{ \ket{b_r} \}$
with equal prior probabilities are respectively $P_\A \coloneqq 3 x_2^2$ and $P_\B \coloneqq 3 y_2^2$
(see Theorem~4 of Ref.~\cite{Eld-2003-unamb});
thus, the average success probability of $\hPi^{(\hA^\opt)}$ is
\begin{eqnarray}
 1 - (1 - P_\A) (1 - P_\B) &=& 3 (x_2^2 + y_2^2 - 3 x_2^2 y_2^2).
\end{eqnarray}
On the other hand, that of an optimal measurement for $\{ \ket{\Psi_r} \}$
with equal prior probabilities is given by
\begin{eqnarray}
 3 \tx_{\upsilon_0}^2 &=&3 \tx_2^2 = 3 \left[ (1 - x_2^2) y_2^2 + x_2^2(1 - 2y_2^2) \right] \nonumber \\
 &=& 3 (x_2^2 + y_2^2 - 3 x_2^2 y_2^2),
\end{eqnarray}
where the first equality follows from the fact that $\tx_0 \ge \tx_2$ (i.e., $\upsilon_0 = 2$)
holds when $y_1 = y_2$,
and the second equality follows from the definition of $\tx_n$.
Thus, $\hPi^{(\hA^\opt)}$ is globally optimal.

We should note that the same discussion is applicable to the case of $x_1 = x_2$
(i.e., $\KA$ is positive real);
in this case, there also exists a globally optimal sequential measurement $\hPi^{(\hA^\opt)}$
satisfying Eq.~\eqref{eq:sym3_optLambda}.

\subsection{Sufficient condition for Theorem~\ref{thm:sym3_optLambda}} \label{subsec:sym3_optLambda_sufficient1}

Since we have already proved the theorem in the case of $y_1 = y_2$,
in what follows, we only consider the case $y_1 \neq y_2$.
(We do not have to assume $x_1 \neq x_2$; the following proof is also valid for $x_1 = x_2$.)
After some algebra using Eq.~\eqref{eq:XG}
\footnote{We also use the fact that 
$p^{(\omega_{1,j})}_j = 1$,
$p^{(\omega_{2,j})}_{j \oplus 1} = p^{(\omega_{2,j})}_{j \oplus 2} = 3\eta$,
and $p^{(\omega_3)}_r = 3y_2^2$.}, we get $\rank~\hGopt(\omega) = 2$
for any $\omega \in \Omega^\opt$ and $\hGopt(\omega) \ket{\pi^\opt_\omega} = 0$,
where $\ket{\pi^\opt_\omega} \in \mHA$ $~(\omega \in \Omega^\opt)$
is the normal vector defined as
\begin{eqnarray}
 \ket{\pi^\opt_\omega} &\coloneqq&
  \left\{
   \begin{array}{cc}
    \displaystyle C_1 \hVA^j \sum_{n=0}^2 x_n^{-1} \ket{\phiA_n}, & \omega = \omega_{1,j}, \\
    \displaystyle C_2 \hVA^j \sum_{n=0}^2 x_n^{-1} z_{\upsilon_n}^{-1} \ket{\phiA_n}, & \omega = \omega_{2,j}, \\
    \ket{\phiA_{\upsilon_2}}, & \omega = \omega_3 \\
   \end{array}
  \right. \label{eq:pi}
\end{eqnarray}
and $C_1$ and $C_2$ are normalization constants.
Thus, it follows that $\hA^\opt$ satisfies Eq.~\eqref{eq:sym3_optLambda}
and Eq.~\eqref{eq:cond} with $\hA = \hA^\opt$ if only if
$\hA^\opt$ is expressed as
\begin{eqnarray}
 \hA^\opt(\omega) &=&
  \left\{
   \begin{array}{cc}
    \kappa^\opt_\omega \ket{\pi^\opt_\omega} \bra{\pi^\opt_\omega}, & \omega \in \Omega^\opt, \\
    0, & {\rm otherwise}, \\
   \end{array}
  \right. \label{eq:Aopt_kappa}
\end{eqnarray}
where, for each $\omega \in \Omega^\opt$, $\kappa^\opt_\omega$ is a nonnegative real number.
Therefore, from Remark~\ref{remark:opt}, to prove that $\hPi^{(\hA^\opt)}$ is an optimal measurement,
it suffices to show that there exists Alice's POVM $\hA^\opt$ (i.e., $\hA^\opt \in \POVMA$)
in the form of Eq.~\eqref{eq:Aopt_kappa}.

Let $\hA$ be an optimal solution to Problem~P.
Due to the symmetry of the states, we assume without loss of generality that
$\hA$ is symmetric in the following sense:
for any $\omega \in \Omega$, $\hA(\omega') = \hVA \hA(\omega) \hVA^\dagger$ and
$\hA(\omega'') = \hVA^\dagger \hA(\omega) \hVA$ hold,
where $\omega', \omega'' \in \Omega$
are the indices such that
$\hB^{(\omega')}_r = \hVB \hBw_{r \ominus 1} \hVB^\dagger$ and 
$\hB^{(\omega'')}_r = \hVB^\dagger \hBw_{r \oplus 1} \hVB$ for any $r \in \mI_3$
(see Theorem~4 of Ref.~\cite{Nak-Kat-Usu-2018-seq_gen} in detail).

Let
\begin{eqnarray}
 \hS(\hT) &\coloneqq&
  \frac{1}{3} \sum_{k=0}^2 \hVA^k \hT \left( \hVA^k \right)^\dagger, \label{eq:Somega}
\end{eqnarray}
where $\hT$ is a positive semidefinite operator on $\mHA$.
$\hS(\hT)$ is also a positive semidefinite operator on $\mHA$
satisfying $\Tr[\hS(\hT)] = \Tr~\hT$ and commuting with $\hVA$.
For notational simplicity, we denote $\hS[\hA(\omega)]$ by $\hS(\omega)$.
Due to the symmetry of $\hA$, $\hS(\omega) = \hS(\omega') = \hS(\omega'')$ holds.
Let
\begin{eqnarray}
 \hE^\opt_k &\coloneqq& \hS \left( \ket{\pi^\opt_{\omega_k}} \bra{\pi^\opt_{\omega_k}} \right), ~ k \in \{ 1,2,3 \};
  \label{eq:Ek}
\end{eqnarray}
then, Eqs.~\eqref{eq:pi} and \eqref{eq:Somega} give
\begin{eqnarray}
 \sum_{j=0}^2 \ket{\pi^\opt_{\omega_{k,j}}} \bra{\pi^\opt_{\omega_{k,j}}} &=& 3 \hE^\opt_k, ~~ k \in \{ 1,2 \},
  \nonumber \\
 \ket{\pi^\opt_{\omega_3}} \bra{\pi^\opt_{\omega_3}} &=& \hE^\opt_3.
  \label{eq:pi_E}
\end{eqnarray}
Here, assume that $\hS(\omega)$ can be expressed as
\begin{eqnarray}
 \hS(\omega) &=& \sum_{k=1}^3 w_{\omega,k} \hE^\opt_k, ~~ \forall \omega \in \Omega_+, \nonumber \\
 w_{\omega,k} &\ge& 0, ~~ \forall \omega \in \Omega_+, k \in \{ 1,2,3 \}, \label{eq:thm_cond}
\end{eqnarray}
where
\begin{eqnarray}
 \Omega_+ &\coloneqq& \{ \omega \in \Omega : \hA(\omega) \neq 0 \}
\end{eqnarray}
and $w_{\omega,k}$ is a weight.
Let us choose
\begin{eqnarray}
 \kappa^\opt_\omega &=&
  \left\{
   \begin{array}{cc}
    \displaystyle \frac{w^\opt_k}{3}, & \omega = \omega_{k,j} ~(k \in \{ 1,2 \}), \\
    w^\opt_3, & \omega = \omega_3, \\
   \end{array}
  \right. \nonumber \\
 w^\opt_k &\coloneqq& \int_{\Omega_+} w_{\omega,k} \dw;
  \label{eq:kappa_omega}
\end{eqnarray}
then, from Eq.~\eqref{eq:Aopt_kappa}, we have
\begin{eqnarray}
 \int_\Omega \hA^\opt(\dw) &=& \sum_{k=1}^3 w^\opt_k \hE^\opt_k
  = \int_{\Omega_+} \sum_{k=1}^3 w_{\omega,k} \hE^\opt_k \dw \nonumber \\
 &=& \int_{\Omega_+} \hS(\dw) = \int_\Omega \hA(\dw) = \identA, \label{eq:Aopt_ident}
\end{eqnarray}
where $\identA$ is the identity operator on $\mHA$.
The first equation follows from Eq.~\eqref{eq:pi_E}.
Equation~\eqref{eq:Aopt_ident} yields $\hA^\opt \in \POVMA$.
Therefore, to prove Theorem~\ref{thm:sym3_optLambda},
it suffices to prove Eq.~\eqref{eq:thm_cond}.

\subsection{Reformulation of Eq.~\eqref{eq:thm_cond}} \label{subsec:sym3_optLambda_sufficient2}

For convenience of analysis, we shall reformulate the sufficient condition
given by Eq.~\eqref{eq:thm_cond}.
For any positive semidefinite operator $\hT \neq 0$,
$s_n(\hT)$ is defined as follows:
\begin{eqnarray}
 s_n(\hT) &\coloneqq& \Braket{\phiA_n | \frac{\hS(\hT)}{\Tr[\hS(\hT)]} | \phiA_n}.
  \label{eq:sn}
\end{eqnarray}
From $\sum_{n=0}^2 \braket{\phiA_n | \hS(\hT) | \phiA_n} = \Tr[\hS(\hT)]$,
$\sum_{n=0}^2 s_n(\hT) = 1$ holds.
Let us consider the following point
\begin{eqnarray}
 s(\hT) &\coloneqq& [s_{\upsilon_1}(\hT), s_{\upsilon_0}(\hT)],
\end{eqnarray}
which is in a two-dimensional space (we call it the $S$-plane).
Since $s_n(\hT) \ge 0$ holds from Eq.~\eqref{eq:sn},
each $s(\hT)$ is in the first quadrant of the $S$-plane.
We can easily verify that the point $s(\hT)$ has a one-to-one correspondence with
$\hS(\hT) / \Tr[\hS(\hT)]$.
Let
\begin{eqnarray}
 e^\opt_k &\coloneqq& s \left( \ket{\pi^\opt_{\omega_k}} \bra{\pi^\opt_{\omega_k}} \right), ~ k \in \{ 1,2,3 \},
  \label{eq:vk}
\end{eqnarray}
which is the point in the $S$-plane that corresponds to $\hE^\opt_k$ defined by Eq.~\eqref{eq:Ek}.
$e^\opt_3 = [0,0]$ holds from Eq.~\eqref{eq:pi}.
Also, let $\mT^\opt$ be the triangle formed by $e^\opt_1$, $e^\opt_2$, and $e^\opt_3$.
Note that $\mT^\opt$ may degenerate to a straight line segment in special cases.
For simplicity, we denote $s_n(\omega) \coloneqq s_n[\hA(\omega)]$
and $s(\omega) \coloneqq s[\hA(\omega)]$ $~(\omega \in \Omega_+)$.
From the first line of Eq.~\eqref{eq:thm_cond}, we have
\begin{eqnarray}
 s(\omega) &=& \frac{1}{\Tr[\hS(\omega)]} \sum_{k=1}^3 w_{\omega,k} e^\opt_k.
\end{eqnarray}
Thus, it follows that Eq.~\eqref{eq:thm_cond} is equivalent to the following:
\begin{eqnarray}
 s(\omega) &\in& \mT^\opt, ~~ \forall \omega \in \Omega_+. \label{eq:thm_sufficient_T}
\end{eqnarray}

\begin{figure}[tb]
 \centering
 \includegraphics[scale=0.8]{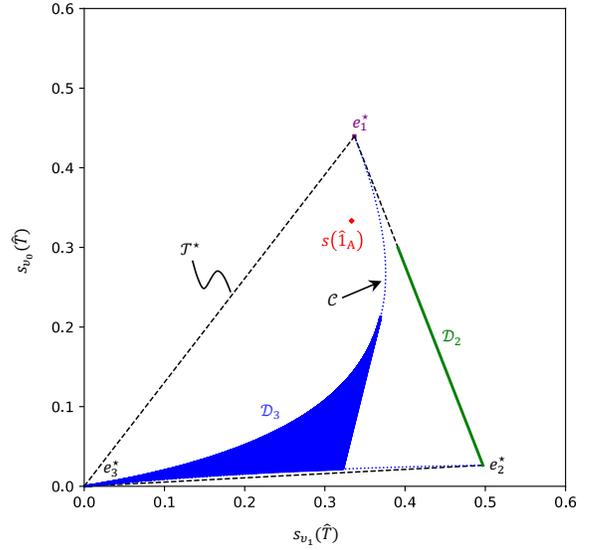}
 \caption{$S$-plane representation in the case of $\KA = \KB = 0.2 \exp(i \pi / 10)$.
 $e^\opt_1$ (purple), $\mD_2$ (green), and $\mD_3$ (blue) respectively show
 the entire sets of points $s(\omega)$ with $|\Tw| = 1$, 2, and 3.
 The dashed line represents the triangle $\mT^\opt$.}
 \label{fig:triangle}
\end{figure}

Figure~\ref{fig:triangle} shows the $S$-plane representation
in the case of $\KA = \KB = 0.2 \exp(i \pi / 10)$.
The entire sets of points $s(\omega)$ $~(\omega \in \Omega_+)$ with $|\Tw| = 2$ and 3,
denoted by $\mD_2$ and $\mD_3$,
are depicted by the green and blue regions in this figure, respectively.
Also, $s(\omega) = e^\opt_1$ holds when $\omega$ satisfies $|\Tw| = 1$.
Indeed, in this case, since $\Tw = \{ j \}$ holds for certain $j \in \mI_3$,
we can easily see that
$\hA(\omega) \propto \ket{\pi^\opt_{\omega_{1,j}}} \bra{\pi^\opt_{\omega_{1,j}}}$
must hold from Eq.~\eqref{eq:suppA}, which gives $s(\omega) = e^\opt_1$.
The triangle $\mT^\opt$ is also shown in the dashed line in Fig.~\ref{fig:triangle}.
One can see that $e^\opt_1$, $\mD_2$, and $\mD_3$ are all included in $\mT^\opt$.
Note that we can show, under the assumption that Theorem~\ref{thm:sym3_optLambda} holds,
that a sequential measurement can be globally optimal if and only if
$s(\identA) (= [1/3, 1/3]) \in \mT^\opt$ holds
(see Appendix~\ref{append:S}).
In the case shown in Fig.~\ref{fig:triangle},
$s(\identA)$ is in $\mT^\opt$, and thus a sequential measurement can be globally optimal.

\subsection{Proof of Eq.~\eqref{eq:thm_sufficient_T}} \label{subsec:sym3_optLambda_proof}

We shall prove that Eq.~\eqref{eq:thm_sufficient_T}, which is a sufficient condition
of Theorem~\ref{thm:sym3_optLambda}, holds.
$\hGopt$ can be rewritten as the following form:
\begin{eqnarray}
 \hGopt(\omega) &=& \hXGopt - \sum_{r=0}^2 \muw_r \ket{a_r} \bra{a_r}. \label{eq:hGw3}
\end{eqnarray}
From Eq.~\eqref{eq:hGw2}, $\muw_r = -\infty$ holds if $\ew_r \neq 0$;
otherwise, $\muw_r = \pw_r / 3$ holds.
It is very hard to show in a naive way that each $s(\omega)$ with $\omega \in \Omega_+$
is included in $\mT^\opt$.
However, we can rather easily show that Eq.~\eqref{eq:thm_sufficient_T} holds
by considering the following two cases:
(1) the case in which at least two of $\{ \muw_r \}_r$ are the same
and (2) the other case in which $\{ \muw_r \}_r$ are all different.

\subsubsection*{Case (1): at least two of $\{ \muw_r \}_r$ are the same}

Due to the symmetry of the states,
we assume $\muw_1 = \muw_2 \eqqcolon q$ without loss of generality; then,
$\hGopt(\omega)$ can be expressed as
\begin{eqnarray}
 \hGopt(\omega) &=& \hXGopt - q \hPsi - p \ket{a_0} \bra{a_0}, \label{eq:K_omega_q}
\end{eqnarray}
where $\hPsi \coloneqq \sum_{k=0}^2 \ket{a_k} \bra{a_k}$ and $p$ is a real number.
If $|\Tw| = 3$, then $q = \pw_1/3 = \pw_2/3$ holds to satisfy Eq.~\eqref{eq:K_omega_q}.
Also, in this case, we can easily see that
$\pw_1 = \pw_2 \le p^{(\omega_2)}_1 = p^{(\omega_2)}_2 = 3\eta$ holds,
which gives $0 \le q \le \eta$.
Moreover, $q = - \infty$ holds if $|\Tw| = 1$, and $q = \eta$ holds if $|\Tw| = 2$.
Thus, $q \le \eta$ always holds.

For each $q \le \eta$, let $\ket{\gamma_q}$ be a normal vector satisfying
\begin{eqnarray}
 \ket{\gamma_q} &\in& \Ker~\hGone(q), \nonumber \\
 \hGone(q) &\coloneqq& \hXGopt - q \hPsi - p_q \ket{a_0} \bra{a_0}, \label{eq:Gone}
\end{eqnarray}
where $p_q$ is a real number determined such that $\rank~\hGone(q) < 3$.
$\ket{\gamma_q}$ can be written, up to a global phase, as (see Appendix~\ref{append:gamma}):
\begin{eqnarray}
 \ket{\gamma_q} &=&
  \left\{
   \begin{array}{cc}
    \displaystyle C'_q \sum_{n=0}^2 \frac{1}{x_n(y_{\upsilon_n}^2 - q)} \ket{\phiA_n}, & q \neq y_2^2, \\
    \ket{\phiA_{\upsilon_2}}, & {\rm otherwise}, \\
   \end{array}
  \right. \label{eq:soq}
\end{eqnarray}
where $C'_q$ is a normalization constant.
Let $\mC$ be the set defined as
\begin{eqnarray}
 \mC &\coloneqq& \{ s(\ket{\gamma_q} \bra{\gamma_q}) : q \le \eta \}.
\end{eqnarray}
In Fig.~\ref{fig:triangle}, $\mC$ is shown in the blue dotted line.
Since $\hGopt(\omega)$ is in the form of Eq.~\eqref{eq:K_omega_q}
satisfying $\rank~\hGopt(\omega) < 3$,
$\hGopt(\omega)$ is equivalent to $\hGone(\muw_1)$.
Thus, $\hA(\omega) \propto \ket{\gamma_{\muw_1}} \bra{\gamma_{\muw_1}}$ holds,
which yields $s(\omega) \in \mC$.
Therefore, to prove $s(\omega) \in \mT^\opt$, it suffices to show 
$\mC \subseteq \mT^\opt$.
We can prove this using Eq.~\eqref{eq:soq} (see Appendix~\ref{append:C}).

\subsubsection*{Case (2): $\{ \muw_r \}_r$ are all different}

In this case, we can show that each $s(\omega)$ is on a straight line segment
whose endpoints are in $\mC$ (see Appendix~\ref{append:case2}).
Since $\mC \subseteq \mT^\opt$ holds, such a line segment is in the triangle $\mT^\opt$.
Therefore, $s(\omega) \in \mT^\opt$ holds.

The two cases (1) and (2) exhaust all possibilities;
thus, from the above arguments, Eq.~\eqref{eq:thm_sufficient_T} holds,
and thus we complete the proof.
\QED

\section{Examples} \label{sec:example}

In this section, we present some examples of symmetric ternary pure states
in which a sequential measurement can be globally optimal.
In Secs.~\ref{subsec:example_bi_eq} and \ref{subsec:example_bi_ineq},
we consider the bipartite case.
In Sec.~\ref{subsec:example_multi}, we consider the multipartite case.

\subsection{Case of $\KA = \KB$} \label{subsec:example_bi_eq}

We first give some examples of bipartite states
$\{ \ket{\Psi_r} \coloneqq \ket{a_r} \otimes \ket{b_r} \}_r$ with $\KA = \KB \eqqcolon K$.
Note that when $\{ \ket{a_r} \}$ and $\{ \ket{b_r} \}$ are in the form of
Eqs.~\eqref{eq:xy_cond} and \eqref{eq:ab_matrix}, $x_n = y_n$ holds for each $n \in \mI_3$,
and thus $x_0 \ge x_1$ holds from $y_0 \ge y_1$.

\begin{figure}[tb]
 \centering
 \includegraphics[scale=0.8]{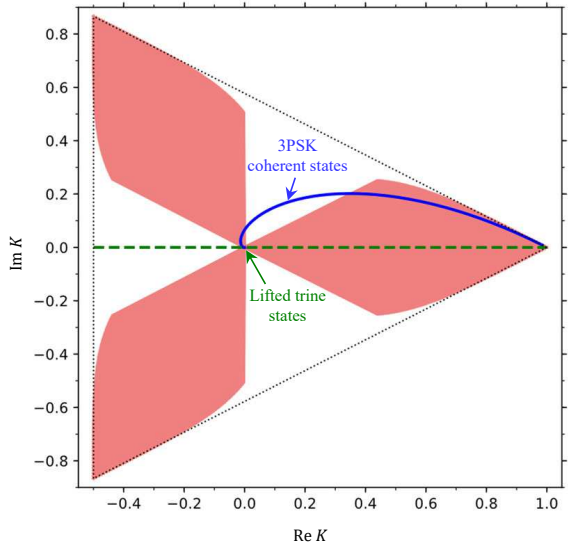}
 \caption{The region of the complex plane where
 a sequential measurement for symmetric ternary pure states with $\KA = \KB \eqqcolon K$
 can be globally optimal.}
 \label{fig:result-half}
\end{figure}

The region of the complex plane where
a sequential measurement for the states $\{ \ket{\Psi_r} \}$ with $\KA = \KB = K$ can be globally optimal
is shown in red in Fig.~\ref{fig:result-half}.
This region is easily obtained from Corollary~\ref{cor:sym3_nas}.
The horizontal and vertical directions are the real and imaginary axes, respectively.
The region of all possible $K$ is represented as the dotted equilateral triangle.
This figure implies that, at least in the case of $\KA = \KB$,
a sequential measurement can be globally optimal in quite a few cases.

As a concrete example, let us consider the symmetric ternary pure states
in which $\{ \ket{a_r} \}$ and $\{ \ket{b_r} \}$ are the lifted trine states,
$\{ \ket{\lift_r} \}_{r=0}^2$, which are expressed by \cite{Sho-2002}
\begin{eqnarray}
 \ket{\lift_r} &=& \sqrt{1 - g}
  \left( \cos \frac{2\pi r}{3} \ket{u_0} + \sin \frac{2\pi r}{3} \ket{u_1} \right)
  + \sqrt{g} \ket{u_2}, \nonumber \\
 \label{eq:lift_r}
\end{eqnarray}
where $\{ \ket{u_n} \}_{n=0}^2$ is an orthonormal basis.
The real parameter $g$ is in the range $0 < g < 1$.
Equation~\eqref{eq:lift_r} gives $K = (3g - 1) / 2$, and thus
$K$ is real in the range $-1/2 < K < 1$.
It follows that the states $\{ \ket{\Psi_r} = \ket{\lift_r} \otimes \ket{\lift_r} \}_r$
are also regarded as lifted trine states.
The region of possible values of $K$ is shown in dashed green line in Fig.~\ref{fig:result-half}.
From this figure, a sequential measurement for $\{ \ket{\Psi_r} \}$ can be globally optimal
if and only if $K \ge 0$ (i.e., $g \ge 1/3$).

\begin{figure}[tb]
 \centering
 \includegraphics[scale=0.8]{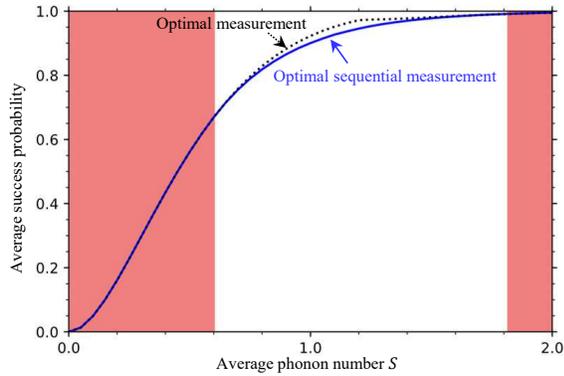}
 \caption{The average success probabilities of an optimal sequential measurement
 and an optimal measurement for the ternary PSK optical coherent states
 $\{ \ket{\alpha_r} \otimes \ket{\alpha_r} \}_r$,
 where $\{ \ket{\alpha_r} \}$ are the ternary PSK optical coherent states
 with an average photon number $S$.}
 \label{fig:result-half-PSK}
\end{figure}

Another example is the states in which $\{ \ket{a_r} \}$ and $\{ \ket{b_r} \}$ are
the ternary PSK optical coherent states $\{ \ket{\alpha_r} \}_{r=0}^2$,
where $\ket{\alpha_r}$ is a normalized eigenvector of the photon annihilation operator
with the eigenvalue $\alpha_r \coloneqq \sqrt{S}\tau^r$,
and $S = |\alpha_r|^2$ is the average photon number of $\ket{\alpha_r}$.
In this case, the states $\{ \ket{\Psi_r} = \ket{\alpha_r} \otimes \ket{\alpha_r} \}_r$ are
also regarded as the ternary PSK optical coherent states with the average photon number $2S$.
We have
\begin{eqnarray}
 K &=& \braket{\alpha_0|\alpha_1} = e^{-\frac{3}{2} S} e^{i \frac{\sqrt{3}}{2} S}
  \label{eq:PSK_K}
\end{eqnarray}
for some global phase.
The solid blue line in Fig.~\ref{fig:result-half} shows the region of possible values of $K$.
It follows from Eq.~\eqref{eq:PSK_K} that $\arg K = \frac{\sqrt{3}}{2} S$ is proportional to $S$.
From Fig.~\ref{fig:result-half},
a necessary and sufficient condition that a sequential measurement for $\{ \ket{\Psi_r} \}$ can be globally optimal
is $2\pi k/3 \le \arg K + \pi/6 \le 2\pi k/3 + \pi/3$, i.e.,
\begin{eqnarray}
 \frac{(4k-1)\pi}{3\sqrt{3}} \le S \le \frac{(4k+1)\pi}{3\sqrt{3}}, ~~ k \in \{ 0, 1, 2, \cdots \}.
  \label{eq:ex_S}
\end{eqnarray}
The average success probability of an optimal sequential measurement for
$\{ \ket{\Psi_r} = \ket{\alpha_r} \otimes \ket{\alpha_r} \}_r$
is plotted in solid blue line in Fig.~\ref{fig:result-half-PSK}.
Also, that of an optimal measurement is shown in dashed black line.
These probabilities can be numerically computed using a modified version of the method
given in Ref.~\cite{Nak-Kat-Usu-2018-Dolinar}.
The region of $S$ satisfying Eq.~\eqref{eq:ex_S},
in which a sequential measurement can be globally optimal, is shown in red.
It is worth mentioning that, as shown in Fig.~2 of Ref.~\cite{Nak-Kat-Usu-2018-Dolinar},
in the strategy for minimum-error discrimination,
an optimal sequential measurement for the ternary PSK optical coherent states
is unlikely to be globally optimal, at least when $S$ is small.
In the strategy for unambiguous discrimination,
a sequential measurement can be globally optimal if (and only if) $S$ satisfies Eq.~\eqref{eq:ex_S}.

\subsection{Case of $\KA \neq \KB$} \label{subsec:example_bi_ineq}

Two concrete examples of symmetric ternary pure states with $\KA \neq \KB$ will be given.
The first is a set of states $\{ \ket{\Psi_r} \coloneqq \ket{a_r} \otimes \ket{\delta_r} \}_r$,
where $\{ \ket{\delta_r} \}_{r=0}^2$ are ternary PPM optical coherent states expressed by
\begin{eqnarray}
 \ket{\delta_0} &\coloneqq& \ket{\alpha} \otimes \ket{\beta} \otimes \ket{\beta}, \nonumber \\
 \ket{\delta_1} &\coloneqq& \ket{\beta} \otimes \ket{\alpha} \otimes \ket{\beta}, \nonumber \\
 \ket{\delta_2} &\coloneqq& \ket{\beta} \otimes \ket{\beta} \otimes \ket{\alpha}.
  \label{eq:PPM}
\end{eqnarray}
$\ket{\alpha}$ and $\ket{\beta}$ are distinct optical coherent states.
$\{ \ket{a_r} \}$ are not necessarily ternary PPM optical coherent states.
From Eq.~\eqref{eq:PPM},
$\KB = \braket{\delta_r | \delta_{r \oplus 1}} = |\braket{\alpha|\beta}|^2 \braket{\beta|\beta}$ holds,
and thus $\KB$ is nonnegative real.
Therefore, as described in Sec.~\ref{subsec:sym3_optLambda_y1y2},
a sequential measurement can be globally optimal.
The same argument can be applied to
states $\{ \ket{\Psi_r} \coloneqq \ket{\delta_r} \otimes \ket{b_r} \}_r$.

\begin{figure}[tb]
 \centering
 \includegraphics[scale=0.8]{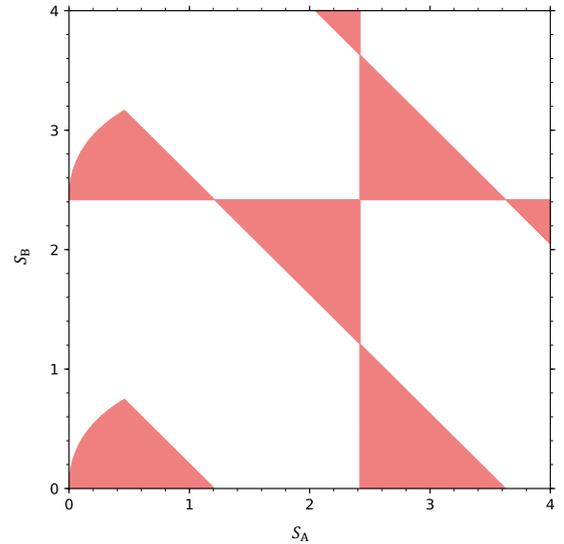}
 \caption{The region where a sequential measurement for the ternary PSK optical coherent states
 $\{ \ket{a_r} \otimes \ket{b_r} \}_r$
 can be globally optimal,
 where $\{ \ket{a_r} \}$ and $\{ \ket{b_r} \}$ are the ternary PSK optical coherent states
 with average photon numbers $\SA$ and $\SB$, respectively.}
 \label{fig:result-PSK}
\end{figure}

The second example is the states $\{ \ket{\Psi_r} \coloneqq \ket{a_r} \otimes \ket{b_r} \}_r$,
where $\{ \ket{a_r} \}$ and $\{ \ket{b_r} \}$ are the ternary PSK optical coherent states
with average photon numbers $\SA$ and $\SB$, respectively.
The states $\{ \ket{\Psi_r} \}$ are also regarded as the ternary PSK optical coherent states
with the average photon number $\SA + \SB$.
The region of $(\SA, \SB)$ in which a sequential measurement can be globally optimal
is shown in red in Fig.~\ref{fig:result-PSK}.
We can see that a sequential measurement can be globally optimal in some cases.
If $\SA$ (or $\SB$) is equal to $4\pi k/(3\sqrt{3}) \approx 2.42k$ $~(k = 1, 2, \cdots)$,
then, since $\KA$ (or $\KB$) is nonnegative real, a sequential measurement can be globally optimal.

Let us consider whether
the ternary PSK optical coherent states $\{ \ket{\alpha_r} \}$ with an average photon number $S$
can be unambiguously discriminated by a Dolinar-like receiver,
which consists of continuous photon counting and infinitely fast feedback (e.g., \cite{Dol-1976}).
The performance of this receiver never exceeds that of an optimal sequential measurement for $N$-partite PSK optical coherent states
$\{ \ket{\alpha'_r}^{\otimes N} \}$ with $N \to \infty$,
where $\{ \ket{\alpha'_r} \coloneqq \ket{\alpha_r / \sqrt{N}} \}_r$ is also the PSK optical coherent states with
the average photon number $S / N$.
Note that $n$ identical copies of $\ket{\alpha'_r}$
are regarded as $\ket{\alpha_r}$ whose average photon number is $S$
(i.e., $\ket{\alpha_r} = \ket{\alpha'_r}^{\otimes N}$).
We here want to know whether a Dolinar-like receiver can be globally optimal.
We consider the bipartite ternary states $\{ \ket{\alpha_r} = \ket{a_r} \otimes \ket{b_r} \}_r$,
where $\ket{a_r} = \ket{\sqrt{t} \alpha_r} (= \ket{\alpha'_r}^{\otimes tN})$ and
$\ket{b_r} = \ket{\sqrt{1-t} \alpha_r} (= \ket{\alpha'_r}^{\otimes (1-t)N})$
with $0 < t < 1$ are optical coherent states with average photon numbers $tS$ and $(1-t)S$,
respectively.
% $tN$ and $(1-t)N$ identical copies of $\ket{\alpha'_r}$
The average success probability of an optimal sequential measurement for
the bipartite states with any $t$ is
an upper bound on that of an optimal sequential measurement for $N$-partite states
$\{ \ket{\alpha'_r}^{\otimes N} \}$ with $N \to \infty$,
and thus is an upper bound on that of a Dolinar-like receiver.
We here show that there exists $t$ such that
an optimal sequential measurement for the corresponding bipartite states
$\{ \ket{a_r} \otimes \ket{b_r} \}_r$ is not globally optimal,
which means that a Dolinar-like receiver cannot be globally optimal.
In the case in which $\braket{\alpha_0 | \alpha_1}$ is nonnegative real
(i.e., $S = 4\pi k/\sqrt{3}$ with $k = 1, 2, \cdots$),
we choose $t = 1/2$; then, from Eq.~\eqref{eq:PSK_K},
$\KA = \KB$ and $\arg~\KA = \pi$ holds,
and thus a sequential measurement cannot be globally optimal,
as already shown in Fig.~\ref{fig:result-half}.
In the other case, we choose $t \to 0$;
formulating $\{ \ket{a_r} \}$ and $\{ \ket{b_r} \}$ in the form of
Eqs.~\eqref{eq:xy_cond} and \eqref{eq:ab_matrix},
we have that for each $k \in \{1,2\}$
\begin{eqnarray}
 x_k^2 &=& \frac{1}{3} + \frac{2}{3} e^{-\frac{3tS}{2}}
  \cos \left[ (-1)^k \frac{2\pi}{3} + \frac{\sqrt{3}tS}{2} \right].
\end{eqnarray}
Taking the limit of $t \to 0$, we obtain $x_2 / x_1 \to 0$.
From Corollary~\ref{cor:sym3_nas},
it is necessary to satisfy $x_2 z_0 - x_1 z_1 \ge 0$ for
a sequential measurement to be able to be globally optimal.
When $t \to 0$, from $x_2 / x_1 \to 0$, $z_1 \to 0$ must hold.
However, $z_1$ converges to a positive number.
($z_1 \to 0$ holds only if $\braket{b_0 | b_1}$ converges to a nonnegative real number,
i.e., $y_1 - y_2 \to 0$;
however, $\braket{b_0 | b_1}$ converges to $\braket{\alpha_0 | \alpha_1}$,
which is not a nonnegative real number.)
Therefore, a Dolinar-like receiver cannot be globally optimal for any ternary PSK optical coherent states.

\subsection{Case of multipartite states} \label{subsec:example_multi}

As an example of multipartite states,
let us address the problem of multiple-copy state discrimination
\cite{Bro-Mei-1996,Aci-Bag-Bai-Mas-2005,Hig-Boo-Dph-Bar-2009,Cal-Vic-Mun-Bag-2010,Hig-Doh-Bar-Pry-Wis-2011}.
We again consider $N$-partite ternary PSK optical coherent states
$\{ \ket{\alpha'_r}^{\otimes N} \}$
$~(\ket{\alpha'_r} \coloneqq \ket{\alpha_r / \sqrt{N}})$.
As described in Sec.~\ref{subsec:example_bi_ineq},
in the limit of $N \to \infty$, a sequential measurement cannot be globally optimal.
In this section, we consider $N$ to be finite.

\begin{figure}[tb]
 \centering
 \includegraphics[scale=0.8]{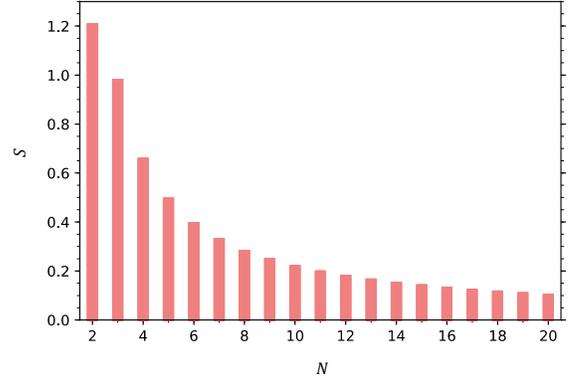}
 \caption{Sufficient condition for a sequential measurement for the $N$-partite
 ternary PSK optical coherent states $\{ \ket{\alpha'_r}^{\otimes N} \}$
 to be able to be globally optimal.
 $S$ is the average photon number of $\ket{\alpha'_r}^{\otimes N}$.
 }
 \label{fig:result-PSK-multiple}
\end{figure}

By using Corollaries~\ref{cor:sym3_nas} and \ref{cor:multipartite},
we can judge whether a sequential measurement can be globally optimal.
The region of the average photon number $S$ of $\ket{\alpha_r}$ for which
the sufficient condition holds
is shown in red in Fig.~\ref{fig:result-PSK-multiple}.
We here consider the range $S \le 1.3$.
We can see in this figure that
a sequential measurement can be globally optimal even for large $N$ (such as $N = 20$)
if $S$ is sufficiently small (such as $S \le 0.1$).

\section{Conclusion} \label{conclusion}

An unambiguous sequential measurement for bipartite symmetric ternary pure states
has been investigated.
We have shown that
a certain type of sequential measurement can always be globally optimal
whenever there exists a globally optimal sequential measurement.
From this result, we have derived a formula that can easily determine
whether an optimal sequential measurement is globally optimal.
Moreover, our results have been extended to multipartite states
and have given a sufficient condition that a sequential measurement can be globally optimal.

\begin{acknowledgments}
 We are grateful to O. Hirota of Tamagawa University for support.
 This work was supported by JSPS KAKENHI Grant Numbers JP17H07115 and JP16H04367.
\end{acknowledgments}

\appendix

\section{Proof of Corollary~\ref{cor:sym3_nas}} \label{append:cor_sym3_nas}

Since, as already described in Sec.~\ref{subsec:sym3_optLambda_y1y2},
a sequential measurement can be globally optimal when $y_1 = y_2$,
we only have to consider the case $y_1 \neq y_2$.

$(1) \Rightarrow (2)$:
From the discussion in Sec.~\ref{subsec:sym3_optLambda_sufficient1},
there exists an optimal solution, $\hA^\opt$, to Problem~P that is expressed by
Eq.~\eqref{eq:Aopt_kappa} with $\kappa^\opt_{\omega_k}$ $~(k \in \{ 1,2,3 \})$ independent of $j \in \mI_3$.
Since $\hA^\opt$ is a POVM, we have
\begin{eqnarray}
 \sum_{j=0}^2 [\hA^\opt(\omega_{1,j}) + \hA^\opt(\omega_{2,j})] + \hA^\opt(\omega_3) &=& \identA.
  \label{eq:Aopt_POVM}
\end{eqnarray}
Substituting Eqs.~\eqref{eq:pi} and \eqref{eq:Aopt_kappa} into Eq.~\eqref{eq:Aopt_POVM} gives
\begin{eqnarray}
 \left[
  \begin{array}{ccc}
   x_{\upsilon_0}^{-2} & x_{\upsilon_0}^{-2} z_0^{-2} & 0 \\
   x_{\upsilon_1}^{-2} & x_{\upsilon_1}^{-2} z_1^{-2} & 0 \\
   x_{\upsilon_2}^{-2} & x_{\upsilon_2}^{-2} z_2^{-2} & 1 \\
  \end{array}
 \right]
 \left[
  \begin{array}{c}
   3 \kappa^\opt_{\omega_1} |C_1|^2 \\ 3 \kappa^\opt_{\omega_2} |C_2|^2
    \\ \kappa^\opt_{\omega_3}
  \end{array}
 \right] &=&
 \left[
  \begin{array}{c}
   1 \\ 1 \\ 1
  \end{array}
 \right], \label{eq:nu2}
\end{eqnarray}
where we use $\upsilon_{\upsilon_k} = k$ $~(k \in \mI_3)$, which follows from Eq.~\eqref{eq:upsilon}.
After some algebra, we can see that Eq.~\eqref{eq:sym3_nas_cond} must hold
if and only if there exists $\kappa^\opt_{\omega_k} \ge 0$ satisfying Eq.~\eqref{eq:nu2}.

$(2) \Rightarrow (1)$:
Let $\kappa^\opt_{\omega_k}$ be the solution to Eq.~\eqref{eq:nu2};
then, $\hA^\opt$ defined by Eq.~\eqref{eq:Aopt_kappa} is a POVM.
Since, as already described in Sec.~\ref{subsec:sym3_optLambda_sufficient1},
$\hGopt(\omega) \ket{\pi^\opt_\omega} = 0$ holds for any $\omega \in \Omega^\opt$,
Eq.~\eqref{eq:cond} with $\hA = \hA^\opt$ obviously holds.
Therefore, from Remark~\ref{remark:opt}, the sequential measurement $\Pi^{(\hA^\opt)}$
is globally optimal.
\QED

Note that one can obtain an analytical expression of $\hA^\opt$
by substituting the solution $\kappa^\opt_{\omega_k}$ to Eq.~\eqref{eq:nu2}
into Eq.~\eqref{eq:pi}.
% However, the expression is rather complicated and we omit it.

\section{Deriving of Eq.~\eqref{eq:ZG}} \label{append:sym3_opt}

From Theorem~4 of Ref.~\cite{Eld-2003-unamb},
since the states $\{ \ket{\Psi_r} \}_{r=0}^2$ with equal prior probabilities
are {\it geometrically uniform} states,
the {\it equal-probability} measurement $\hPi^\opt \coloneqq \{ \hPi^\opt_r \}_{r=0}^3$,
given by
\begin{eqnarray}
 \hPi^\opt_r &=& \ket{\pi^\opt_r} \bra{\pi^\opt_r}, ~~ r \in \mI_3, \nonumber \\
 \ket{\pi^\opt_r} &\coloneqq& \frac{\tx_{\upsilon_0}}{\sqrt{3}} \sum_{n=0}^2 \tx_n^{-1} \ket{\tphi_n},
\end{eqnarray}
is an optimal unambiguous measurement for $\{ \ket{\Psi_r} \}$.
Also, its average success probability is $P(\hPi^\opt) = 3 \tx_{\upsilon_0}^2$.
Therefore, $\hZG^\opt$ of Eq.~\eqref{eq:ZG}, which is a feasible solution to Problem~$\DPG$,
satisfies $\Tr~\hZG^\opt = P(\hPi^\opt)$,
and thus is an optimal solution to Problem~$\DPG$.
Note that $\hZG^\opt$ of Eq.~\eqref{eq:ZG} is always an optimal solution to Problem~$\DPG$,
while there could be other optimal solutions.

\section{Supplement of the $S$-plane} \label{append:S}

Under the assumption that Theorem~\ref{thm:sym3_optLambda} holds, we shall show
that $s(\identA) \in \mT^\opt$ is a necessary and sufficient condition
that a sequential measurement can be globally optimal.

First, we show the necessity.
Assume that a sequential measurement can be globally optimal.
From Theorem~\ref{thm:sym3_optLambda}, there exists $\hA^\opt \in \POVMA$ satisfying
Eq.~\eqref{eq:sym3_optLambda} such that $\hPi^{(\hA^\opt)}$ is globally optimal.
As described in Sec.~\ref{subsec:sym3_optLambda_sufficient1},
$\hA^\opt$ is expressed by Eq.~\eqref{eq:Aopt_kappa}.
Thus, let $\kappa'_3 \coloneqq \kappa^\opt_{\omega_3}$ and
$\kappa'_k \coloneqq 3 \kappa^\opt_{\omega_k}$ for $k \in \{1,2\}$;
then, since $\hA^\opt$ is a POVM, we have
\begin{eqnarray}
 \sum_{k=1}^3 \kappa'_k \hE^\opt_k &=& \int_\Omega \hA^\opt(\dw) = \identA. \label{eq:Asum}
\end{eqnarray}
Premultiplying and postmultiplying this equation by $\bra{\phiA_n}$ and $\ket{\phiA_n}$, respectively,
gives
\begin{eqnarray}
 \sum_{k=1}^3 \frac{\kappa'_k}{3} e^\opt_k &=& s(\identA).
  \label{eq:sn_A}
\end{eqnarray}
This indicates that $s(\identA)$ is the weighted sum of $e^\opt_k$ with the weights
$\kappa'_k/3 \ge 0$,
and thus $s(\identA) \in \mT^\opt$ holds.

Next, we show the sufficiency.
The above argument can be applied in the reverse direction.
Assume $s(\identA) \in \mT^\opt$; then, there exists $\kappa'_k \ge 0$ satisfying
Eq.~\eqref{eq:sn_A}.
Consider $\hA^\opt$ expressed by Eq.~\eqref{eq:Aopt_kappa} with
$\kappa^\opt_{\omega_3} = \kappa'_3$ and $\kappa^\opt_{\omega_k} = \kappa'_k / 3$ $~(k = \{1,2\})$.
It follows that $\hA^\opt$ is a POVM satisfying Eq.~\eqref{eq:sym3_optLambda} and
$\hGopt(\omega) \hA^\opt(\omega) = 0$.
Thus, from Remark~\ref{remark:opt}, $\hPi^{(\hA^\opt)}$ is globally optimal,
and thus a sequential measurement can be globally optimal.
\QED

\section{Supplement of Theorem~\ref{thm:sym3_optLambda}}

\subsection{Proof of $y_2^2 < \eta < y_1^2$} \label{append:eta}

Let
\begin{eqnarray}
 \chi &\coloneqq& y_0^2y_1^2 + y_1^2y_2^2 + y_2^2y_0^2; \label{eq:chi}
\end{eqnarray}
then, we have
\begin{eqnarray}
 \lefteqn{ (1-3y_k^2)^2 - (1-3\chi) } \nonumber \\
 &=& 3 (3 y_k^4 - 2y_k^2 + \chi) \nonumber \\
 &=& 3 [y_k^4 - 2 y_k^2 (y_{k \oplus 1}^2 + y_{k \oplus 2}^2) + y_k^2 (y_{k \oplus 1}^2 + y_{k \oplus 2}^2)
  + y_{k \oplus 1}^2 y_{k \oplus 2}^2] \nonumber \\
 &=& 3 (y_k^2 - y_{k \oplus 1}^2)(y_k^2 - y_{k \oplus 2}^2), \label{eq:eta_ineq}
\end{eqnarray}
where the third line follows from $\sum_{n=0}^2 y_n^2 = 1$.
Substituting $k = 1$ into Eq.~\eqref{eq:eta_ineq} yields $(1-3y_1^2)^2 \le 1-3\chi$.
The equality holds when $y_0 = y_1$.
In this case, from $y_2^2 = 1 - 2y_0^2$, we have $1 - 3y_1^2 = y_2^2 - y_0^2 < 0 \le \sqrt{1 - 3\chi}$.
Thus, $1 - 3y_1^2 < \sqrt{1 - 3\chi}$ always holds.
Substituting the definition of $y_n$ in Eq.~\eqref{eq:xy} into Eq.~\eqref{eq:chi} gives
$|\KB|^2 = |\KB'|^2 = 1 - 3 \chi$.
Therefore, from the definition of $\eta$, we have
\begin{eqnarray}
 \eta &=& \frac{1}{3} \left( 1 - \sqrt{1 - 3\chi} \right) < y_1^2. \label{eq:eta_chi}
\end{eqnarray}
In the same way, substituting $k = 2$ into Eq.~\eqref{eq:eta_ineq} yields $1-3y_2^2 > \sqrt{1-3\chi}$,
which gives $\eta > y_2^2$.
\QED

\subsection{Derivation of $\ket{\gamma_q}$} \label{append:gamma}

From Eqs.~\eqref{eq:ab_matrix} and \eqref{eq:XG}, we have
\begin{eqnarray}
 \hXGopt - q \hPsi &=& 3 \sum_{n=0}^2 x_n^2 (y_{\upsilon_n}^2 - q) \ket{\phiA_n} \bra{\phiA_n}.
  \label{eq:X_qPsi}
\end{eqnarray}
Also, since $y_2^2 < \eta < y_1^2 \le y_0^2$ holds (see Appendix~\ref{append:eta}),
$\hXGopt - q \hPsi$ $~(q \le \eta)$ is singular if and only if $q = y_2^2$ holds.
One can easily see $\ket{\gamma_q} = \ket{\phiA_{\upsilon_2}}$ when $q = y_2^2$.
Note that, in this case, one can define $p_q \coloneqq 0$.

In what follows, assume $q \neq y_2^2$.
From Eq.~\eqref{eq:Gone}, we have
\begin{eqnarray}
 (\hXGopt - q \hPsi) \ket{\gamma_q} &=& p_q \ket{a_0} \braket{a_0 | \gamma_q} \propto \ket{a_0}. \label{eq:soq0}
\end{eqnarray}
Thus, from Eqs.~\eqref{eq:ab_matrix} and \eqref{eq:X_qPsi}, we have
\begin{eqnarray}
 \ket{\gamma_q} &\propto& (\hXGopt - q \hPsi)^{-1} \ket{a_0}
  \propto \sum_{n=0}^2 \frac{1}{x_n(y_{\upsilon_n}^2 - q)} \ket{\phiA_n}.
  \label{eq:soq1}
\end{eqnarray}
Therefore, $\ket{\gamma_q}$ is expressed by Eq.~\eqref{eq:soq}.
One can verify $\rank~\hGone(q) < 3$ by letting
$p_q \coloneqq \braket{a_0 | (\hXGopt - q\Psi)^{-1} | a_0}^{-1}$ if $q < \eta$
and $p_q \coloneqq - \infty$ if $q = \eta$.
Note that since $\ket{\gamma_q}$ is unique up to a global phase,
$\rank~\hGone(q) = 2$ holds.

\subsection{Proof of $\mC \subseteq \mT^\opt$} \label{append:C}

Since $s(\ket{\gamma_q}\bra{\gamma_q}) = e^\opt_3 \in \mT^\opt$ holds when $q = y_2^2$,
we have only to consider the case of $q \neq y_2^2$.
Let
\begin{eqnarray}
 u_n(q) &\coloneqq& (y_n^2 - q)^{-1}. \label{eq:uk}
\end{eqnarray}
Note that $q \neq y_n^2$ holds for any $n \in \mI_3$ since
$y_2^2 < \eta < y_1^2$ holds (see Appendix~\ref{append:eta}).
From Eqs.~\eqref{eq:Somega}, \eqref{eq:sn}, and \eqref{eq:soq} and $\upsilon_{\upsilon_n} = n$,
we have
\begin{eqnarray}
 s_{\upsilon_n}(\ket{\gamma_q}\bra{\gamma_q}) &\propto&
  x_{\upsilon_n}^{-2} u_n^2(q), \label{eq:snwq}
\end{eqnarray}
and thus
\begin{eqnarray}
 s(\ket{\gamma_q}\bra{\gamma_q}) &\propto& [x_{\upsilon_1}^{-2}u_1^2(q), x_{\upsilon_0}^{-2}u_0^2(q)].
  \label{eq:sgg}
\end{eqnarray}

First, let us consider the case in which
the three points $e^\opt_1$, $e^\opt_2$, and $e^\opt_3$ lie on a straight line.
From Eq.~\eqref{eq:pi}, this case occurs only when $y_0 = y_1$.
Since $s(\ket{\gamma_q}\bra{\gamma_q}) \propto [x_{\upsilon_1}^{-2}, x_{\upsilon_0}^{-2}]$
holds from Eq.~\eqref{eq:sgg},
every point in $\mC$ is on the line joining the origin $e^\opt_3$ to the point $e^\opt_1$.
From $u_0^2(q) = u_1^2(q) \le u_2^2(q)$ (see Appendix~\ref{append:uk}),
we have
\begin{eqnarray}
 s_{\upsilon_0}(\ket{\gamma_q} \bra{\gamma_q})
  &=& \frac{x_{\upsilon_0}^{-2} u_0^2(q)}{\displaystyle \sum_{k=0}^2 x_{\upsilon_k}^{-2} u_k^2(q)}
  \le \frac{x_{\upsilon_0}^{-2} u_0^2(q)}{\displaystyle u_0^2(q) \sum_{k=0}^2 x_{\upsilon_k}^{-2}}
  = s_{\upsilon_0}(\omega_1). \nonumber \\
\end{eqnarray}
Thus, $s(\ket{\gamma_q} \bra{\gamma_q})$ is an interior point between $e^\opt_1$ and $e^\opt_3$.
Therefore, $\mC \subseteq \mT^\opt$ holds.

Next, let us consider the other case in which
$e^\opt_1$, $e^\opt_2$, and $e^\opt_3$ do not lie on a straight line.
Let $l_{jk}$ denote the straight line joining $e^\opt_j$ and $e^\opt_k$.
It suffices to prove the following two statements:
(a) $\mC$ is in the region between the two lines $l_{13}$ and $l_{23}$,
and (b) $\mC$ is in the region between the two lines $l_{12}$ and $l_{13}$.

First, we prove the statement (a).
The gradient of the line joining the origin to the point $s(\ket{\gamma_q}\bra{\gamma_q})$ is
\begin{eqnarray}
 \zeta(q) &\coloneqq&
  \frac{s_{\upsilon_0}(\ket{\gamma_q}\bra{\gamma_q})}{s_{\upsilon_1}(\ket{\gamma_q}\bra{\gamma_q})}
 = \frac{x_{\upsilon_1}^2(y_1^2 - q)^2}{x_{\upsilon_0}^2(y_0^2 - q)^2},
\end{eqnarray}
where the last equality follows from Eq.~\eqref{eq:snwq}.
Since $\eta < y_1^2 \le y_0^2$ holds (see Appendix~\ref{append:eta}),
one can easily verify that $\zeta(q)$ monotonically decreases in the range $q \le \eta$,
which gives
\begin{eqnarray}
 \zeta(-\infty) &\ge& \zeta(q) \ge \zeta(\eta), ~~ \forall q \le \eta. \label{eq:zeta_dec}
\end{eqnarray}
Also, from $e^\opt_1 = s(\ket{\gamma_{-\infty}}\bra{\gamma_{-\infty}})$ and
$e^\opt_2 = s(\ket{\gamma_\eta}\bra{\gamma_\eta})$,
the gradients of the lines $l_{13}$ and $l_{23}$ are, respectively, $\zeta(-\infty)$ and $\zeta(\eta)$.
Therefore, from Eq.~\eqref{eq:zeta_dec}, the statement (a) holds.

Next, we prove the statement (b).
Let $c(q)$ denote the $s_{\upsilon_1}$-coordinate of the intersection of the $s_{\upsilon_1}$-axis
and the line joining the two points $e^\opt_1$ and $s(\ket{\gamma_q}\bra{\gamma_q})$ in $\mC$.
It follows that the statement (b) holds if and only if $c(q)$ satisfies
\begin{eqnarray}
 0 &\le& c(q) \le c(\eta), ~~ \forall q \le \eta. \label{eq:cq}
\end{eqnarray}
Since $s(\ket{\gamma_q}\bra{\gamma_q})$ is on the line joining $e^\opt_1$ and $[c(q), 0]$,
we have that for some real number $w$
\begin{eqnarray}
 s_{\upsilon_1}(\ket{\gamma_q}\bra{\gamma_q}) &=& w s_{\upsilon_1}(\omega_1) + (1 - w) c(q), \nonumber \\
 s_{\upsilon_0}(\ket{\gamma_q}\bra{\gamma_q}) &=& w s_{\upsilon_0}(\omega_1). \label{eq:case1_c1}
\end{eqnarray}
Also, since $\sum_{n=0}^2 s_n(\hT) = 1$ holds for any nonzero positive semidefinite operator $\hT$,
we have
\begin{eqnarray}
 s_{\upsilon_2}(\ket{\gamma_q}\bra{\gamma_q}) &=& w s_{\upsilon_2}(\omega_1) + (1 - w) [1 - c(q)].
  \label{eq:case1_c2}
\end{eqnarray}
After some algebra with Eqs.~\eqref{eq:case1_c1}, \eqref{eq:case1_c2}, and \eqref{eq:snwq},
we obtain
\begin{eqnarray}
 \tc(q) &\coloneqq& \frac{x_{\upsilon_1}^2 c(q)}{x_{\upsilon_2}^2 [1 - c(q)]}
  = \frac{u_1^2(q) - u_0^2(q)}{u_2^2(q) - u_0^2(q)}. \label{eq:tc}
\end{eqnarray}
It follows from the definition of $\tc(q)$ that $\tc(q)$ monotonically increases with $c(q)$.
Thus, the statement (b), i.e. Eq.~\eqref{eq:cq}, is equivalent to
\begin{eqnarray}
 0 &\le& \tc(q) \le \tc(\eta), ~~\forall q \le \eta. \label{eq:tcq_ineq}
\end{eqnarray}
Since $u_0^2(q) \le u_1^2(q) \le u_2^2(q)$ holds (see Appendix~\ref{append:uk}),
$\tc(q) \ge 0$ obviously holds.
Therefore, we need only show $\tc(q) \le \tc(\eta)$.

Differentiating $\tc(q)$ of Eq.~\eqref{eq:tc} with respect to $q$ gives
\begin{eqnarray}
 \frac{d\tc(q)}{dq} &=& \frac{2[u_1^2(q) - u_0^2(q)]}{u_2^2(q) - u_0^2(q)} [f[u_1(q)] - f[u_2(q)]], \nonumber \\
 f(x) &\coloneqq& \frac{x^2+u_0(q)x+u_0^2(q)}{x+u_0(q)},
\end{eqnarray}
which implies that $d\tc(q)/dq \ge 0$ is equivalent to $f[u_1(q)] \ge f[u_2(q)]$.
In the case of $q < y_2^2$, from $0 \le u_1(q) \le u_2(q)$,
$f[u_1(q)] \le f[u_2(q)]$ (i.e., $d\tc(q)/dq \le 0$) holds, which follows from
the fact that $f(x)$ monotonically increases in the range $x \ge 0$.
In the other case of $q > y_2^2$,
from $u_2(q) \le - u_1(q) \le - u_0(q)$ (see Appendix~\ref{append:uk}),
$f[u_2(q)] < 0 \le f[u_1(q)]$ (i.e., $d\tc(q)/dq \ge 0$) holds, which follows from
the fact that $f(x) < 0$ holds if and only if $x < -u_0(q)$.
Therefore, $\tc(q)$ $~(q \le \eta)$ attains its maximum at $q = -\infty$ and/or $q = \eta$,
and thus, for the rest, it suffices to show $\tc(-\infty) \le \tc(\eta)$.

From Eq.~\eqref{eq:tc}, we have
\begin{eqnarray}
 \tc(q) &=& \frac{(y_2^2-q)^2 [ (y_0^2-q)^2 - (y_1^2-q)^2 ]}
  {(y_1^2-q)^2 [ (y_0^2-q)^2 - (y_2^2-q)^2 ]} \nonumber \\
 &=& \frac{(y_0^2-y_1^2)(y_2^2-q)^2(y_0^2+y_1^2-2q)}{(y_0^2-y_2^2)(y_1^2-q)^2(y_0^2+y_2^2-2q)}, \label{eq:tcq}
\end{eqnarray}
which gives
\begin{eqnarray}
 \tc(-\infty) &=& \frac{y_0^2 - y_1^2}{y_0^2 - y_2^2}.
\end{eqnarray}
Also, we have
\begin{eqnarray}
 \lefteqn{ (y_1^2-\eta)^2(y_0^2+y_2^2-2\eta) - (y_2^2-\eta)^2(y_0^2+y_1^2-2\eta) } \nonumber \\
 &=& (y_1^2-y_2^2)(y_0^2y_1^2 + y_1^2y_2^2 + y_2^2y_0^2 - 2\eta + 3\eta^2) \nonumber \\
 &=& (y_1^2-y_2^2)(3\eta^2 - 2\eta + \chi) \nonumber \\
 &=& 0,
\end{eqnarray}
where the second to fourth lines, respectively, follow from $\sum_{k=0}^2 y_k^2 = 1$,
Eq.~\eqref{eq:chi}, and Eq.~\eqref{eq:eta_chi}.
Thus, substituting $q = \eta$ into Eq.~\eqref{eq:tcq} gives
$\tc(\eta) = (y_0^2 - y_1^2) / (y_0^2 - y_2^2)$.
Therefore, $\tc(-\infty) = \tc(\eta)$ holds.
\QED

\subsection{Proof of $u_0^2(q) \le u_1^2(q) \le u_2^2(q)$ $~(\forall q \le \eta, q \neq y_2^2)$} \label{append:uk}

Since $u_1(q) \ge u_0(q) > 0$ holds, it suffices to prove $u_2^2(q) \ge u_1^2(q)$.
In the case of $q < y_2^2$, from $u_2(q) > u_1(q) > 0$, this is obvious.
Let us consider the case of $q > y_2^2$.
Since $u_2(q) < 0$ holds, it suffices to show $u_2(q) + u_1(q) \le 0$.
Let $\tu_k \coloneqq u_k(\eta)$; then, we have
\begin{eqnarray}
 \tu_2 + \tu_1 &\le& \tu_2 + \tu_1 + \tu_0 \nonumber \\
 &=& \tu_2\tu_1\tu_0 [(\tu_0\tu_1)^{-1} + (\tu_1\tu_2)^{-1} + (\tu_2\tu_0)^{-1}] \nonumber \\
 &=& \tu_2\tu_1\tu_0 (3\eta^2 - 2\eta + \chi) \nonumber \\
 &=& 0,
\end{eqnarray}
where the third and fourth lines, respectively, follow from the definition of $\chi$ in Eq.~\eqref{eq:chi}
and Eq.~\eqref{eq:eta_chi}.
Since $u_2(q) \le \tu_2$ and $u_1(q) \le \tu_1$ hold,
$u_2(q) + u_1(q) \le \tu_2 + \tu_1 \le 0$ holds.
\QED

\subsection{Case (2)} \label{append:case2}

Let us consider, without loss of generality, $\omega \in \Omega_+$ such that
$\muw_2 < \muw_0$ and $\muw_2 < \muw_1$.
In order to show that $s(\omega)$ is on a straight line segment whose endpoints are in $\mC$,
we shall show the two statements:
(a) $s(\omega)$ is on a certain straight line segment,
and (b) the line segment is part of a straight line segment whose endpoints are in $\mC$.

Since we now consider the case (2), $|\Tw|$ must be 2 or 3.
Let
\begin{eqnarray}
 \hX &\coloneqq&
  \left\{
   \begin{array}{cc}
    \hXGopt - \muw_2 \Psi, & |\Tw| = 3, \\
    \hXGopt + \infty \ket{a_2}\bra{a_2}, & |\Tw| = 2. \\
   \end{array}
  \right. \label{eq:X}
\end{eqnarray}
One can easily see that $\hX$ is a positive definite operator.
Let $\ket{\varpi(q)}$ be a normal vector satisfying
\begin{eqnarray}
 \ket{\varpi(q)} &\in& \Ker~\hGtwo(q), \nonumber \\
 \braket{a_0|\varpi(q)} &\ge& 0, \nonumber \\
 \hGtwo(q) &\coloneqq& \hX - q \ket{a_1} \bra{a_1} - p'_q \ket{a_0} \bra{a_0}, \label{eq:hGw_q2}
\end{eqnarray}
where $p'_q$ is the function of $q$ such that $\rank~\hGtwo(q) < 3$.
(We can define such $p'_q$ as
$p'_q \coloneqq \braket{a_0 | (\hX - q \ket{a_1} \bra{a_1})^{-1} | a_0}^{-1}$.
Since $\hX - q \ket{a_1} \bra{a_1}$ is positive definite, such $p'_q$ always exists.)
$p'_q$ monotonically decreases with $q$.
$\hGopt(\omega) = \hGtwo[\muw_1 - \muw_2]$ holds if $|\Tw| = 3$;
otherwise, $\hGopt(\omega) = \hGtwo[\muw_1]$ holds.

First, we show the statement (a).
Let $\hGtwo_0 \coloneqq \hGtwo(0)$, $\hGtwo_1 \coloneqq \hGtwo(p'_0)$,
$\ket{\varpi_0} \coloneqq \ket{\varpi(0)}$,
and $\ket{\varpi_1} \coloneqq \ket{\varpi(p'_0)}$.
Note that $p'_q = 0$ holds when $q = p'_0$ (i.e., $p'_{p'_0} = 0$).
We shall express $\ket{\varpi(q)}$ in terms of $\ket{\varpi_0}$ and $\ket{\varpi_1}$.
For each $k \in \{0,1\}$, from $\hGtwo_k \ket{\varpi_k} = 0$, we have
\begin{eqnarray}
 \ket{a_k} &=& \frac{\hX \ket{\varpi_k}}{p'_0 \braket{a_k|\varpi_k}}.
  \label{eq:aXpi}
\end{eqnarray}
Note that since $\hX$ is positive definite, we have $\hX \ket{\varpi_k} \neq 0$,
which yields $p'_0 \braket{a_k|\varpi_k} \neq 0$.

Substituting Eq.~\eqref{eq:hGw_q2} into $\hGtwo(q) \ket{\varpi(q)} = 0$ and
using Eq.~\eqref{eq:aXpi} yields
\begin{eqnarray}
 \ket{\varpi(q)} &=& \hX^{-1} (p'_q r_0 \ket{a_0} + q r_1 \ket{a_1}) \nonumber \\
 &=& \frac{1}{p'_0} \left( \frac{p'_q r_0}{\braket{a_0|\varpi_0}} \ket{\varpi_0}
                     + \frac{q r_1}{\braket{a_1|\varpi_1}} \ket{\varpi_1} \right),
 \label{eq:QSD_kernel_s_sk0}
\end{eqnarray}
where $r_k \coloneqq \braket{a_k|\varpi(q)}$.
Premultiplying this equation by $\bra{a_0}$ and some algebra gives
\begin{eqnarray}
 \frac{q r_1}{\braket{a_1|\varpi_1}} &=& \frac{(p'_0 - p'_q) r_0}{\braket{a_0|\varpi_1}}.
\end{eqnarray}
Substituting this equation into Eq.~\eqref{eq:QSD_kernel_s_sk0} gives
\begin{eqnarray}
 \ket{\varpi(q)} &=& \frac{r_0}{p'_0}
  \left( \frac{p'_q}{\braket{a_0|\varpi_0}} \ket{\varpi_0}
   + \frac{p'_0 - p'_q}{\braket{a_0|\varpi_1}} \ket{\varpi_1} \right). \label{eq:case2_sw}
\end{eqnarray}
Since $r_0 \ge 0$ and $\braket{a_0|\varpi_k} \ge 0$ hold from Eq.~\eqref{eq:hGw_q2},
it follows from Eq.~\eqref{eq:case2_sw} that $\ket{\varpi(q)}$ is expressed as
\begin{eqnarray}
 \ket{\varpi(q)} &=& c_0 \ket{\varpi_0} + c_1 \ket{\varpi_1} \label{eq:case2_c01}
\end{eqnarray}
with certain nonnegative real numbers $c_0$ and $c_1$.
Let $q_2$ be the real number satisfying $p'_{q_2} = q_2$.
One can easily verify that, when $q = q_2$,
Eq.~\eqref{eq:case2_c01} with $c_0 = c_1 \eqqcolon c$ holds.
Let $\ket{\varpi_2} \coloneqq \ket{\varpi(q_2)}$.

Due to the symmetry of the states,
$\hS(\ket{\varpi_0} \bra{\varpi_0}) = \hS(\ket{\varpi_1} \bra{\varpi_1}) \eqqcolon \hS_{\varpi_0}$ holds.
Thus, from Eq.~\eqref{eq:case2_c01}, we have
\begin{eqnarray}
 \hS(\ket{\varpi(q)} \bra{\varpi(q)})
  &=& (c_0^2 + c_1^2) \hS_{\varpi_0} + c_0 c_1 \hS',
  \label{eq:QSD_R_Rs}
\end{eqnarray}
where
\begin{eqnarray}
 \hS' &\coloneqq& \frac{1}{3} \sum_{j=0}^2 \hVA^j
  \left( \ket{\varpi_0} \bra{\varpi_1} + \ket{\varpi_1} \bra{\varpi_0} \right)
  \left( \hVA^j \right)^\dagger.
\end{eqnarray}
Substituting $q = q_2$ into Eq.~\eqref{eq:QSD_R_Rs} and letting
$\hS_{\varpi_2} \coloneqq \hS(\ket{\varpi_2} \bra{\varpi_2})$ yields
\begin{eqnarray}
 \hS_{\varpi_2} &=& 2 c^2 \hS_{\varpi_0} + c^2 \hS'.
\end{eqnarray}
Substituting this into Eq.~\eqref{eq:QSD_R_Rs} gives
\begin{eqnarray}
 \hS(\ket{\varpi(q)} \bra{\varpi(q)}) &=& c'_0 \hS_{\varpi_0} + c'_2 \hS_{\varpi_2},
  \label{eq:case2_Sw}
\end{eqnarray}
where $c'_0 \coloneqq (c_0 - c_1)^2$ and $c'_2 = c_0 c_1 / c^2$.
Note that taking the trace of this gives $c'_0 + c'_2 = 1$
and that $c'_0, c'_2 \ge 0$ holds.
Also, Eq.~\eqref{eq:case2_Sw} gives
\begin{eqnarray}
 s(\ket{\varpi(q)} \bra{\varpi(q)}) &=& c'_0 s_{\varpi_0} + c'_2 s_{\varpi_2},
\end{eqnarray}
where $s_{\varpi_k} \coloneqq s(\ket{\varpi_k} \bra{\varpi_k})$ for each $k \in \{0,2\}$.
Therefore, $s(\ket{\varpi(q)} \bra{\varpi(q)})$ is on the straight line segment, denoted by $\mL$,
whose endpoints are $s_{\varpi_0}$ and $s_{\varpi_2}$.

Next, we show the statement (b).
In the case of $q = q_2$, since Eq.~\eqref{eq:hGw_q2} with $q = p'_q = q_2$ holds,
this is the case (1), i.e., at least two of $\{ \muw_r \}_r$ in Eq.~\eqref{eq:hGw3} are the same;
thus, $s_{\varpi_2}$ is in $\mC$.
Also, if $\omega$ satisfies $|\Tw| = 3$, then
$q = 0$ is also the case (1), and thus $s_{\varpi_0}$ is in $\mC$.
Therefore, in the case of $|\Tw| = 3$,
$\mL$ is the line segment whose endpoints, $s_{\varpi_0}$ and $s_{\varpi_2}$, are in $\mC$.
In what follows, assume $|\Tw| = 2$.
We shall show that $\mL$ is part of the straight line segment whose endpoints are
$e^\opt_1 = s(\omega_1) \in \mC$ and $s_{\varpi_2} \in \mC$.
% As mentioned earlier, $e^\opt_1$ is in $\mC$.
Taking the limit as $q \to - \infty$ in Eq.~\eqref{eq:hGw_q2} gives
$s(\ket{\varpi(-\infty)} \bra{\varpi(-\infty)}) = e^\opt_1$.
Thus, repeating the above argument with $q \to - \infty$ indicates that
$\ket{\varpi(-\infty)}$ is expressed as Eq.~\eqref{eq:case2_c01} with
$c_0 > 0$ and $c_1 < 0$,
and that Eq.~\eqref{eq:case2_Sw} holds with $c'_2 < 0$.
Thus, $s_{\varpi_0}$ is an interior point between $e^\opt_1$ and $s_{\varpi_2}$.
Therefore, $\mL$ is part of the line segment whose endpoints are $e^\opt_1$ and $s_{\varpi_2}$.
\QED

%\bibliography{quant}

\end{document}